\documentstyle[12pt,fleqn,epsf,epsfig,axodraw]{article}
\renewcommand{\theequation}{\arabic{equation}}

\def\lsim{\raise0.3ex\hbox{$\;<$\kern-0.75em\raise-1.1ex\hbox{$\sim\;$}}}
\def\gsim{\raise0.3ex\hbox{$\;>$\kern-0.75em\raise-1.1ex\hbox{$\sim\;$}}}

\begin{document}
\setlength{\unitlength}{1cm}
\setlength{\mathindent}{0cm}
\thispagestyle{empty}
\null
\hfill WUE-ITP-03-014\\
\null
\hfill UWThPh-2003-19\\
\null
\hfill HEPHY-PUB 777/03\\
\null
\hfill hep-ph/0308141\\
\vskip .8cm
\begin{center}
	{\Large \bf CP asymmetries in neutralino production
		         in $e^+e^-$-collisions}
\vskip 2.5em
{\large
{\sc A.~Bartl$^{a}$\footnote{e-mail:
        bartl@ap.univie.ac.at}
     H.~Fraas$^{b}$\footnote{e-mail:
        fraas@physik.uni-wuerzburg.de},
	  O.~Kittel$^{a,b}$\footnote{e-mail:
		  kittel@physik.uni-wuerzburg.de},
	  W.~Majerotto$^{c}$\footnote{e-mail:
        majer@qhepu3.oeaw.ac.at}
}}\\[1ex]
{\normalsize \it
$^{a}$ Institut f\"ur Theoretische Physik, Universit\"at Wien, 
Boltzmanngasse 5, A-1090 Wien, Austria}\\
{\normalsize \it
$^{b}$ Institut f\"ur Theoretische Physik, Universit\"at
W\"urzburg, Am Hubland, D-97074~W\"urzburg, Germany}\\
{\normalsize \it
$^{c}$ Institut f\"ur Hochenergiephysik, \"Osterreichische
Akademie der Wissenschaften, Nikolsdorfergasse 18, 
A-1050 Wien, Austria}\\
\vskip 1em
\end{center} \par
\vskip .8cm

\begin{abstract}

We study two CP sensitive triple-product asymmetries for
neutralino production 
$e^+\,e^- \to\tilde{\chi}^0_i \, \tilde{\chi}^0_j$
and the subsequent leptonic two-body decay 
$\tilde{\chi}^0_i \to \tilde{\ell} \, \ell$, 
$ \tilde{\ell} \to \tilde{\chi}^0_1 \, \ell$, for
$ \ell= e,\mu,\tau$.
We calculate the asymmetries, cross sections and branching ratios
in the Minimal Supersymmetric Standard Model with 
complex parameters $\mu$ and $M_1$.
We present numerical results for the asymmetries to be expected
at a linear electron-positron collider in the 500 GeV range.
The asymmetries can go up to 25\%.
We estimate the event rates which are necessary to 
observe the asymmetries. Polarized electron and positron beams 
can significantly enhance the asymmetries and cross sections.
In addition, we show how the two
decay leptons can be distinguished by making use of their energy
distributions.

\end{abstract}

\newpage

\section{Introduction}

In the Standard Model there is only one physical CP-violating phase in the 
Cabibbo-Kobayashi-Maskawa  matrix. The 
Minimal Supersymmetric Standard Model (MSSM) \cite{haberkane} 
contains several  new sources of CP violation if the parameters of the model
are complex. 
In the neutralino sector of the MSSM these are the $U(1)$ and $SU(2)$
gaugino mass parameters $M_1$ and $M_2$, respectively, and the higgsino mass
parameter $\mu$. One of these parameters, usually $M_2$, can be made 
real by redefining the fields. 
The non-vanishing phases of $M_1$ and $\mu$ cause CP-violating
effects already at tree level, which could be large and thus  
observable in high energy collider experiments \cite{TDR}.

In this paper we study neutralino production
(for recent studies with complex parameters and polarized beams see
\cite{choi1,kali,gudi1}):
\begin{eqnarray}
	e^+ + e^-&\to& \tilde{\chi}^0_i+\tilde{\chi}^0_j
  \label{production} 
\end{eqnarray}
with longitudinally polarized beams and the subsequent leptonic 
two-body decay of one of the neutralinos
\begin{eqnarray}
   \tilde{\chi}^0_i&\to& \tilde{\ell} + \ell_1, 
  \label{decay_1} 
\end{eqnarray}
and that of the decay slepton
\begin{eqnarray}
  \tilde{\ell}&\to&\tilde{\chi}^0_1+\ell_2; \;\;\; \ell= e,\mu,\tau.
  \label{decay_2}
\end{eqnarray}

T-odd observables \cite{choi1,donoghue,oshimo,valencia,staudecay}
are a useful tool to study the influence of the 
CP-violating parameters $M_1$ and $\mu$.
For the neutralino production (\ref{production}) and  the 
two-body decay chain of the neutralino (\ref{decay_1}) and (\ref{decay_2})
we introduce the triple-product 
 \begin{eqnarray}\label{AT1}
	 {\mathcal T}_{I} &=& (\vec p_{e^-} \times \vec p_{\chi_i}) 
	 \cdot \vec p_{\ell_1},
 \end{eqnarray}
and define the corresponding T-odd asymmetry
 \begin{eqnarray}
	 {\mathcal A}_{I} = \frac{\sigma({\mathcal T}_{I}>0)
						 -\sigma({\mathcal T}_{I}<0)}
							{\sigma({\mathcal T}_{I}>0)+
							\sigma({\mathcal T}_{I}<0)},
\label{TasymmetryI}
\end{eqnarray}
where $\sigma$ is the cross section (\ref{crossection}) for reactions
(\ref{production})-(\ref{decay_2}). Since under time reversal the 
triple-product changes sign, the asymmetry is a T-odd asymmetry.
With the leptonic two-body decay of the slepton Eq.~(\ref{decay_2}),
we can construct a further T-odd observable which does not require the 
identification of the neutralino momentum. We  replace the 
neutralino momentum $\vec p_{\chi_i}$ in Eq.~(\ref{AT1}) by the lepton momentum
$\vec p_{\ell_2}$ from the slepton decay, which defines the triple-product: 
 \begin{eqnarray}
{\mathcal T}_{II} &=& (\vec p_{e^-} \times \vec p_{\ell_2}) 
\cdot \vec p_{\ell_1}
	\label{AT2}
\end{eqnarray}
and the T-odd asymmetry
\begin{eqnarray}
	 {\mathcal A}_{II} = \frac{\sigma({\mathcal T}_{II}>0)
						 -\sigma({\mathcal T}_{II}<0)}
							{\sigma({\mathcal T}_{II}>0)+
							\sigma({\mathcal T}_{II}<0)}.
\label{TasymmetryII}
\end{eqnarray}
Due to CPT invariance these T-odd asymmetries are CP-odd 
if the widths of the exchanged particles and final state interactions 
are neglected, which is done in this work.

The T-odd observables in the production of 
neutralinos at  tree level are due to spin effects. 
Only if there are CP-violating phases in the neutralino
sector and  if two different neutralinos are
%Only if $M_1$ and/or $\mu$ attain non-vanishing physical phases and 
produced, each of the produced neutralinos 
has a polarization vector with a component perpendicular to the production 
plane  \cite{choi1,kali,gudi1}. This polarization  leads to
asymmetries in the angular distributions of the decay products, 
as defined in Eq.~(\ref{TasymmetryI}) and (\ref{TasymmetryII}). 
In general also spin-spin correlations of the neutralinos
are present \cite{gudi1}.
However, we will not study these in the present work.

In Section \ref{Definitions and Formalism} we present the formalism 
used. In Section \ref{T-odd asymmetry} we discuss the qualitative 
properties of the asymmetries. 
We present detailed numerical results in Section \ref{Numerical results}.
Section \ref{Summary and conclusion} contains a short summary and 
conclusion.

\section{Definitions and formalism
  \label{Definitions and Formalism}}

\begin{figure}[h]
%\fbox{
\begin{picture}(5,6.5)(-2,.5)
		\put(1,4.7){$\vec p_{\chi_j}$}
   \put(3.4,6){$\vec p_{e^- }$}
   \put(3,3.8){$\theta $}
   \put(3.3,2.3){$\vec  p_{e^+}$}
   \put(4.8,4.7){$\vec p_{\chi_i}$}
   \put(7.2,5.7){$ \vec p_{\ell_1}$}
	\put(6.6,3.9){$ \theta_{D_1}$}
	\put(7.2,4.6){$ \theta_{1}$}
   \put(6.2,3.){$ \vec p_{\tilde{ \ell} }$}
   \put(9.2,2.3){$\vec  p_{\ell_2}$}
   \put(6,1.7){$ \vec p_{\chi_1}$}
	\put(8.15,1.75){$\theta_{D_2} $}
	\put(8.5,4.3){$ \phi_{D_1}$}
	\put(9.,4.8){$ \phi_{1}$}
	\put(9.8,1.3){$\phi_{2} $}
\end{picture}
\scalebox{1.9}{
\begin{picture}(0,0)(1.3,-0.25)
\ArrowLine(40,50)(0,50)
\Vertex(40,50){2}
\ArrowArcn(40,50)(14,245,180)
\ArrowLine(55,80)(40,50)
\ArrowLine(25,20)(40,50)
\ArrowLine(40,50)(80,50)
\ArrowLine(80,50)(110,75)
\DashLine(80,50)(115,50){1.5}
\ArrowArc(115,50)(6, 65,255)
\ArrowArc(115,50)(12,65,130)
\ArrowArc(80,50)(22,0,40)
\ArrowArcn(80,50)(17,0,305)
\DashLine(80,50)(100,20){4}
\Vertex(80,50){2}
\ArrowLine(100,20)(125,15)
\ArrowArc(100,20)(20,310,346)
\ArrowLine(100,20)(85,0)
\DashLine(100,20)(115,0){1.5}
\ArrowArc(115,0)(16,15,55)
\Vertex(100,20){2}
\end{picture}}
\caption{\label{shematic picture}
Definition of angles and momenta of the production
and decay process.}
\end{figure}
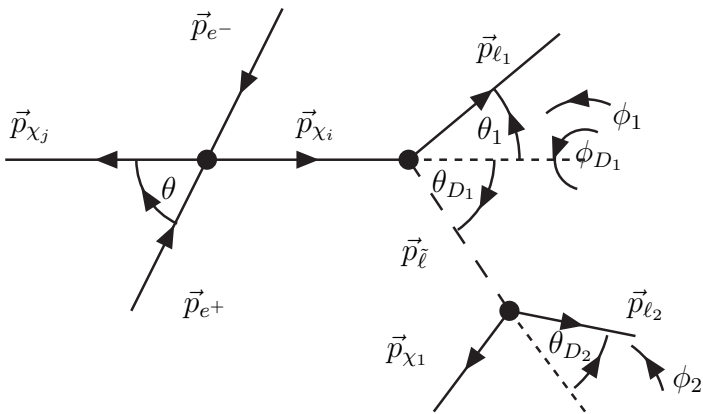

In this section we give the interaction Lagrangians, the complex
couplings and the formulae for the cross section for neutralino 
production (\ref{production}) and decay
(\ref{decay_1}),(\ref{decay_2}).
For the definition of the angles for production and decay,
see Fig.~\ref{shematic picture}.

\subsection{Lagrangian and couplings
     \label{Lagrangian and couplings}}

The interaction Lagrangians for the processes 
(\ref{production})-(\ref{decay_2}) are (in our notation and
conventions we follow closely \cite{haberkane,gudi1}):
\begin{eqnarray}
& & {\cal L}_{Z^0\tilde{\chi}^0_i\tilde{\chi}^0_j} =
\frac{1}{2}\frac{g}{\cos\theta_W}Z_{\mu}\bar{\tilde{\chi}}^0_i\gamma^{\mu}
[O_{ij}^{''L} P_L+O_{ij}^{''R} P_R]\tilde{\chi}^0_j,\\
& & {\cal L}_{\ell \tilde{\ell}\tilde{\chi}^0_i} =
g f^L_{\ell i}\bar{\ell}P_R\tilde{\chi}^0_i\tilde{\ell}_L+
g f^R_{\ell i}\bar{\ell}P_L\tilde{\chi}^0_i\tilde{\ell}_R+\mbox{h.c.},
\quad \ell=e,\mu, \quad i, j=1,\dots,4, \\
& & {\cal L}_{Z^0 \ell^{+} \ell^{-}} =
-\frac{g}{\cos\theta_W}Z_{\mu}\bar{\ell}\gamma^{\mu}[L_{\ell}P_L+
 R_{\ell}P_R]\ell.
\end{eqnarray}
In the neutralino basis $\tilde{\gamma},
\tilde{Z}, \tilde{H}^0_a, \tilde{H}^0_b$ the couplings are:
\begin{eqnarray}
f_{\ell i}^L &=& -\sqrt{2}\bigg[\frac{1}{\cos
\theta_W}(T_{3\ell}-e_{\ell}\sin^2\theta_W)N_{i2}+e_{\ell}\sin \theta_W
N_{i1}\bigg],\nonumber\\
f_{\ell i}^R &=& -\sqrt{2}e_{\ell} \sin \theta_W
\Big[\tan \theta_W N_{i2}^*- N_{i1}^*\Big],
\label{eq_6}\\
O_{ij}^{''L}&=&-\frac{1}{2}\Big(N_{i3}N_{j3}^*-N_{i4}N_{j4}^*\Big)\cos2\beta
  -\frac{1}{2}\Big(N_{i3}N_{j4}^*+N_{i4}N_{j3}^*\Big)\sin2\beta,\nonumber\\
O_{ij}^{''R}&=&-O_{ij}^{''L*},\mbox{ with } i\mbox{,
}j=1,\ldots,4.\label{eq_7}\\
L_{\ell}&=&T_{3\ell}-e_{\ell}\sin^2\theta_W, \quad
 R_{\ell}=-e_{\ell}\sin^2\theta_W \label{eq_5}.
\end{eqnarray}
where $P_{L, R}=\frac{1}{2}(1\mp \gamma_5)$, $g$ is the weak coupling
constant ($g=e/\sin\theta_W$, $e>0$), and $e_\ell$ and $T_{3 \ell}$ denote the
charge and the third component of the weak isospin of the lepton
$\ell$, $\tan \beta=v_2/v_1$ is the ratio of
 the vacuum expectation values of the two neutral Higgs fields.
$N_{ij}$ is the complex unitary $4\times 4$ matrix which diagonalizes
the neutral gaugino-higgsino mass matrix $Y_{\alpha\beta}$, 
 $N_{i \alpha}^*Y_{\alpha\beta}N_{\beta k}^{\dagger}=
 m_{\tilde{\chi}^0_i}\delta_{ik}$.

For the neutralino decay into staus 
$\tilde \chi^0_i \to \tilde \tau_k \tau$, we take stau mixing into
account and write for the Lagrangian \cite{Bartl:2002bh}:
\begin{eqnarray}
& & {\mathcal L}_{\tau\tilde{\tau} \chi_i }=  g\tilde \tau_k \bar \tau
(a^{\tilde \tau}_{ki} P_R+b^{\tilde \tau}_{ki} P_L)\chi^0_i + {\rm h.c.}~,
 \quad k = 1,2; \; i=1,\dots,4, \label{eq:LagStauchi} 
\end{eqnarray}
with
\begin{equation}
a_{kj}^{\tilde \tau}=
({\mathcal R}^{\tilde \tau}_{kn})^{\ast}{\mathcal A}^\tau_{jn},\qquad 
b_{kj}^{\tilde \tau}=
({\mathcal R}^{\tilde \tau}_{kn})^{\ast}
{\mathcal B}^\tau_{jn},\qquad
(n=L,R)
\end{equation}
\begin{equation}
{\mathcal A}^{\tau}_j=\left(\begin{array}{ccc}
f^{L}_{\tau j}\\[2mm]
h^{R}_{\tau j} \end{array}\right),\qquad 
{\mathcal B}^{\tau}_j=\left(\begin{array}{ccc}
h^{L}_{\tau j}\\[2mm]
f^{R}_{\tau j} \end{array}\right),
\label{eq:coupl2}
\end{equation}
with ${\mathcal R}^{\tilde \tau}_{kn}$ the stau mixing 
matrix defined below and 
\begin{eqnarray}
h^{L}_{\tau j}&=& (h^{R}_{\tau j})^{\ast}
=-Y_{\tau}( N_{j3}^{\ast}\cos\beta+N_{j4}^{\ast}\sin\beta), \\ 
Y_{\tau}&=& m_{\tau}/(\sqrt{2}m_W \cos\beta), 
\end{eqnarray}
with $m_W$ the mass of the $W$ boson and
$m_{\tau}$ the mass of the $\tau$-lepton.
The masses and couplings of the
$\tau$-sleptons follow from the 
hermitian $2 \times 2$  $\tilde\tau_L - \tilde \tau_R$ mixing matrix:
\begin{equation}
{\mathcal{L}}_M^{\tilde \tau}= -(\tilde \tau_L^{\ast},\, \tilde \tau_R^{\ast})
\left(\begin{array}{ccc}
M_{\tilde \tau_{LL}}^2 & e^{-i\varphi_{\tilde \tau}}|M_{\tilde \tau_{LR}}^2|\\[5mm]
e^{i\varphi_{\tilde \tau}}|M_{\tilde \tau_{LR}}^2| & M_{\tilde \tau_{RR}}^2
\end{array}\right)\left(
\begin{array}{ccc}
\tilde \tau_L\\[5mm]
\tilde \tau_R \end{array}\right),
\label{eq:mm}
\end{equation}
with
\begin{eqnarray}
M_{\tilde \tau_{LL}}^2 & = & m_{\tilde\ell_L}^2 + m_{\tau}^2 ,\\[3mm]
M_{\tilde \tau_{RR}}^2 & = & m_{\tilde\ell_R}^2 + m_{\tau}^2 ,\\[3mm]
M_{\tilde \tau_{RL}}^2 & = & (M_{\tilde\tau_{LR}}^2)^{\ast}=
  m_{\tau}(A_{\tau}-\mu^{\ast}  
 \tan\beta), \label{eq:mlr}
\end{eqnarray}
\begin{equation}
\varphi_{\tilde \tau}  = \arg\lbrack A_{\tau}-\mu^{\ast}\tan\beta\rbrack ,
\label{eq:phtau}
\end{equation}
$A_{\tau}$ is the complex trilinear scalar coupling parameter.
The $\tilde \tau$ mass eigenstates are 
$(\tilde\tau_1, \tilde \tau_2)=(\tilde \tau_L, \tilde \tau_R)
{{\mathcal R}^{\tilde \tau}}^{T}$ with 
 \begin{equation}
{\mathcal R}^{\tilde \tau}
	 =\left( \begin{array}{ccc}
e^{i\varphi_{\tilde \tau}}\cos\theta_{\tilde \tau} & 
\sin\theta_{\tilde \tau}\\[5mm]
-\sin\theta_{\tilde \tau} & 
e^{-i\varphi_{\tilde \tau}}\cos\theta_{\tilde \tau}
\end{array}\right),
\label{eq:rtau}
\end{equation}
and
\begin{equation}
\cos\theta_{\tilde \tau}=
\frac{-|M_{\tilde \tau_{LR}}^2|}{\sqrt{|M_{\tilde \tau _{LR}}^2|^2+
(m_{\tilde \tau_1}^2-M_{\tilde \tau_{LL}}^2)^2}},\quad
\sin\theta_{\tilde \tau}=\frac{M_{\tilde \tau_{LL}}^2-m_{\tilde \tau_1}^2}
{\sqrt{|M_{\tilde \tau_{LR}}^2|^2+(m_{\tilde \tau_1}^2-M_{\tilde \tau_{LL}}^2)^2}}.
\label{eq:thtau}
\end{equation}
The mass eigenvalues are
\begin{equation}
 m_{\tilde \tau_{1,2}}^2 = \frac{1}{2}\left((M_{\tilde \tau_{LL}}^2+M_{\tilde \tau_{RR}}^2)\mp 
\sqrt{(M_{\tilde \tau_{LL}}^2 - M_{\tilde \tau_{RR}}^2)^2 +4|M_{\tilde \tau_{LR}}^2|^2}\right).
\label{eq:m12}
\end{equation}
In order to reduce the number of MSSM parameters,
we assume the renormalization group
equations (RGE) for the slepton masses \cite{hall}:
\begin{eqnarray}\label{RGE1}
m_{\tilde\ell_R }^2 &=& m_0^2 +0.23 M_2^2
-m_Z^2\cos 2 \beta \sin^2 \theta_W, \\
m_{\tilde\ell_L  }^2 &=& m_0^2 +0.79 M_2^2
+m_Z^2\cos 2 \beta(-\frac{1}{2}+ \sin^2 \theta_W),\label{RGE2}
\end{eqnarray}
where 
$m_Z$ is the mass of the $Z$ boson and 
$m_{0}$ is the scalar mass parameter.

\subsection{Cross sections
     \label{Cross sections}}

In order to calculate the amplitude squared
for the complete process of neutralino production (\ref{production})
and the two-body decay chain of the  neutralino $ \tilde\chi^0_i$ 
(\ref{decay_1})-(\ref{decay_2}), we use the spin density matrix 
formalism of \cite{gudi1,spin}. The amplitude squared can be 
written as:
\begin{eqnarray}       \label{amplitude1}
	  |T|^2 &=& 2~\sum_{\lambda_i \lambda_i'}|\Delta(\tilde{\chi}^{0}_i)|^2  
	  ~|\Delta(\tilde{\ell})|^2
      \rho_P(\tilde{\chi}^{0}_i)^{\lambda_i \lambda_i'}
		\rho_{D1}(\tilde{\chi}^{0}_i)_{\lambda_i'\lambda_i}\;
		D_2(\tilde{\ell}),
\end{eqnarray}
with $\rho_P(\tilde{\chi}^{0}_i)$  the spin density production matrix 
of neutralino $ \tilde{\chi}^{0}_i$, the propagator 
$ \Delta(\tilde{\chi}^0_i)=1/[s_{\chi_i}-m_{\chi_i}^2
	+im_{\chi_i}\Gamma{\chi_i}]$,
the decay matrix 
$\rho_{D1}(\tilde{\chi}^{0}_i)$ for decay (\ref{decay_1}),
the slepton propagator 
$ \Delta(\tilde \ell)=1/[s_{\tilde \ell}-m_{\tilde \ell}^2 +im_{\tilde
\ell}\Gamma_{\tilde \ell}]$,
and the factor $D_2(\tilde \ell)$ for the slepton decay (\ref{decay_2}).
In Eq.~(\ref{amplitude1}), $\lambda_i$ and $\lambda_i'$ are the helicities of the neutralino 
$\tilde{\chi}^0_i$ and 
$s_{\chi_i}=p_{\chi_i}^2(s_{\tilde \ell}=p_{\tilde \ell}^2)$,
the mass and width of 
$\tilde{\chi}^0_i~(\tilde \ell)$ are denoted by
$m_{\chi_i}~(m_{\tilde \ell})$ and $\Gamma_{\chi_i}~(\Gamma_{\tilde \ell})$, 
respectively.
The factor 2 is due to the summation of the helicities of the second
neutralino  $\tilde{\chi}^0_j$, whose decay is not considered.

The spin basis vectors
$s^a_{\chi_i}\;(a=1,2,3)$ 
of the neutralino $ \tilde \chi^0_i$ fulfill the orthonormality relations 
$s^a_{\chi_i}\cdot s^b_{\chi_i}=-\delta^{ab}$ and
$s^a_{\chi_i}\cdot p_{\chi_i}=0$.
Their explicit form is given in Appendix 
\ref{Representation of momentum and spin vectors}.
The (unnormalized) density matrices can then be 
expanded in terms of the Pauli matrices:
   \begin{eqnarray} 
      \rho_P(\tilde{\chi}^{0}_i)^{\lambda_i \lambda_i'} & = &
           \delta_{\lambda_i \lambda_i'} P + 
            \sum_{a} \sigma^{a}_{\lambda_i \lambda_i'}
				\Sigma_P^a,\label{rhoP} \\
                     \label{eq_3}
		\rho_{D1}(\tilde{\chi}^{0}_i)_{\lambda_i' \lambda_i} & = & 
           \delta_{\lambda_i' \lambda_i} D_1 +
			  \sum_a \sigma^a_{\lambda_i'
				  \lambda_i}\Sigma^a_{D1}.\label{rhoD1}
    \end{eqnarray}
With our choice of the spin vectors, $\frac{\Sigma^{3}_P}{P}$
is the longitudinal polarization of neutralino $ \tilde \chi^0_i$,
$\frac{\Sigma^{1}_P}{P}$ is the transverse polarization in the 
production plane and $\frac{\Sigma^{2}_P}{P}$ is the polarization
perpendicular to the production plane.
The analytical formulae
for $P,D_1,D_2$ and $\Sigma^{2}_P,\Sigma^{a}_{D1}$ are given
in Appendix \ref{Neutralino production and decay matrices}.
Inserting the density matrices (\ref{rhoP}) and (\ref{rhoD1})
in Eq.~(\ref{amplitude1}) leads to:
   \begin{eqnarray} \label{amplitude2}
		|T|^2 &=& 4~|\Delta(\tilde{\chi}^{0}_i)|^2~  
		|\Delta(\tilde{\ell})|^2
	  ( P D_1 + \vec \Sigma_P \vec\Sigma_{D1} )\;D_2.
				\end{eqnarray}
The cross section and distributions
in the laboratory system are then obtained by integrating 
$|T|^2$ over the Lorentz invariant phase space element 
$d{\rm Lips}(s,p_{\chi_j },p_{{\ell}_1},p_{\chi_1},p_{{\ell}_2})$
(\ref{Lips}):
\begin{equation}\label{crossection}
		d\sigma=\frac{1}{2 s}|T|^2 
%		(2\pi)^4 \delta^4(p_1+p_2-\sum_{i=3}^8 p_i) 
d{\rm Lips}(s,p_{\chi_j },p_{{\ell}_1},p_{\chi_1},p_{{\ell}_2})\label{eq_13},
		\end{equation}
where we use the narrow width approximation for the propagators.
		
The contributions of the spin correlation terms 
$\vec \Sigma_P \vec\Sigma_{D1} $ 
 to the total cross section vanish.
Their contributions to the energy distributions of the lepton
$\ell_1$ and $\ell_2$ from decay (\ref{decay_1}) and (\ref{decay_2})
vanish due to the Majorana properties of the neutralinos 
\cite{pectovgudi} if CP is conserved. If  CP is violated, they  vanish 
to leading order perturbation theory \cite{pectovgudi}.
In our case, the contributions can be neglected because they are
proportional to the widths of the exchanged particles.

\section{T-odd asymmetry
	\label{T-odd asymmetry}}

The asymmetry ${\mathcal A}_{I}$, defined in Eq.~(\ref{TasymmetryI}), 
can be written in terms of the angular 
distribution of the decay lepton  $\ell_1$:
\begin{eqnarray}
  {\mathcal A}_{I} 
	 =\frac{\int^{0}_{1}   \frac{d\sigma}{d\cos\theta_{I}} d\cos\theta_{I} 
            -\int^{-1}_{0}\frac{d\sigma}{d\cos\theta_{I}} d\cos\theta_{I}}
          {\int^{0}_{1}   \frac{d\sigma}{d\cos\theta_{I}} d\cos\theta_{I}
				 +\int^{-1}_{0}\frac{d\sigma}{d\cos\theta_{I}} d\cos\theta_{I}}
		= \frac{N_+ - N_- }{N_+ + N_-},
 \end{eqnarray}
where $\cos\theta_{I}=\frac{\vec p_{e^-} \times \vec p_{\chi_i}}
{|\vec p_{e^-} \times \vec p_{\chi_i}|} 
\cdot \frac{\vec p_{\ell_1}}{|\vec p_{\ell_1}|}$ 
and 
%{\bf
thus ${\mathcal A}_{I}$ is the difference of the number of 
events with lepton $\ell_1$ above $(N_+)$ and below  $(N_-)$ 
the production plane, normalized by the total number of events. 
%}
In order to measure the asymmetry ${\mathcal A}_{I}$, the production
plane and thus the momentum
$\vec p_{\chi_i}$ of neutralino $\tilde{\chi}^0_i$ has to be
reconstructed. 
We will discuss in Section~\ref{Energy distributions of the leptons}
how to distinguish the two leptons. 

Analogously the asymmetry ${\mathcal A}_{II}$, defined in
Eq.~(\ref{TasymmetryII}), can be written in terms of the angular distribution 
of the decay leptons  $\ell_1$ and $\ell_2$:
\begin{eqnarray}
{\mathcal A}_{II} 
=\frac{\int^{0}_{ 1}\frac{d\sigma}{d\cos\theta_{II}} d\cos\theta_{II} 
		-\int^{-1}_{0}\frac{d\sigma}{d\cos\theta_{II}} d\cos\theta_{II}}
      {\int^{0}_{ 1}\frac{d\sigma}{d\cos\theta_{II}} d\cos\theta_{II} 
		+\int^{-1}_{0}\frac{d\sigma}{d\cos\theta_{II}} d\cos\theta_{II}},
\end{eqnarray}
where $\cos\theta_{II}=\frac{\vec p_{e^-} \times \vec p_{\ell_2}}
{|\vec p_{e^-} \times \vec p_{\ell_2}|} 
\cdot \frac{\vec p_{\ell_1}}{|\vec p_{\ell_1}|}$.
Inserting the cross section (\ref{crossection}) in the definitions of the
asymmetries (\ref{TasymmetryI}) and (\ref{TasymmetryII}) we obtain: 
 \begin{eqnarray}
	 {\mathcal A}_{I,II} 
	 = \frac{\int {\rm Sign}[{\mathcal T}_{I,II}]
		 |T|^2 d{\rm Lips}}
           {\int |T|^2 d{\rm Lips}}
	=  \frac{\int {\rm Sign}[{\mathcal T}_{I,II}]
	\Sigma_P^2 \Sigma_{D1}^2 d{\rm Lips}}
          {\int  P D_1 d{\rm Lips}},
\end{eqnarray}
with
$ d{\rm Lips} =d{\rm Lips}(s,p_{\chi_i},p_{\chi_j} )
             d{\rm Lips}(s_{\chi_i},p_{\tilde {\ell}},p_{{\ell}_1})
				 d{\rm Lips}(s_{\tilde {\ell}},p_{\chi_1},p_{{\ell}_2})
				 \delta(s_{\chi_i}-m_{\chi_i}^2)
				 \delta(s_{\tilde\ell}-m_{\tilde\ell}^2)$,
see Eq.~(\ref{Lips}).  
In the numerator only the spin correlation terms perpendicular to the 
production plane $\Sigma_P^2 \Sigma_{D1}^2$ remain, since only
$\Sigma_P^2 $ contains the triple-products (\ref{AT1}) or (\ref{AT2}).
Thus, the contributions to ${\mathcal A}_{I,II}$ directly stem 
from $\Sigma_P^2 $.

In  case the neutralino decays into a scalar tau,
we take stau mixing into account and the asymmetries are reduced
due to their qualitative dependence on the 
$\tilde \chi^0_i$-$\tilde \tau_k$-$\tau$ couplings:
\begin{eqnarray}\label{Amixing}
	 {\mathcal A}_{I,II}\propto
	 \frac{|a^{\tilde \tau}_{ki}|^2-|b^{\tilde \tau}_{ki}|^2}
	 {|a^{\tilde \tau}_{ki}|^2+|b^{\tilde \tau}_{ki}|^2},
\end{eqnarray}
which can be seen from the expressions of $D_1$~(\ref{plusterm})
and $\Sigma_{D1}^2 $~(\ref{minusterm}). Because the asymmetry
is only proportional to the absolute values of 
$a^{\tilde \tau}_{ki},b^{\tilde \tau}_{ki}$,
it is not sensitive to a possible phase $\varphi_{A_{\tau}}$ of $A_{\tau}$.
In order to be sensitive to $\varphi_{A_{\tau}}$, one would have to consider 
an asymmetry which involves the transverse polarization of the $\tau$.

\section{Numerical results
	\label{Numerical results}}

In the following numerical analysis we study for $\sqrt{s} = 500$
and longitudinally polarized beams with $P_-=0.8$ and $P_+=-0.6$, 
the dependence of the neutralino production cross sections 
$\sigma (e^+e^-\to\tilde{\chi}^0_i\tilde{\chi}^0_j)$, 
the branching ratios $BR(\tilde{\chi}^0_i\to \tilde{\ell} \ell)$ 
and the asymmetries 
${\mathcal A}_I$ and ${\mathcal A}_{II}$ on the parameters
$\mu = |\mu| \, e^{ i\,\varphi_{\mu}}$, 
$M_1 = |M_1| \, e^{ i\,\varphi_{M_1}}$ and $M_2$
 for $\tan \beta=10$.
In order to reduce the number of parameters, we assume 
$|M_1|=5/3 \tan^2\theta_W M_2$ and in Eqs.~(\ref{RGE1}) and 
(\ref{RGE2}) we take $m_0=100$ GeV for the slepton masses.
Since the pair production of equal  neutralinos is not CP sensitive,
we discuss the lightest unequal pairs
$\tilde{\chi}^0_1 \,\tilde{\chi}^0_2$,
$\tilde{\chi}^0_1 \,\tilde{\chi}^0_3$  and 
$\tilde{\chi}^0_2 \,\tilde{\chi}^0_3$.

\subsection{Production of $\tilde\chi^0_1 \, \tilde\chi^0_2$ }

\begin{figure}[h]
 \begin{picture}(20,20)(0,-2)
	\put(2.5,16.5){\fbox{$\sigma(e^+e^- \to\tilde{\chi}^0_1 
			\tilde{\chi}^0_2)$ in fb}}
	\put(0,9){\includegraphics{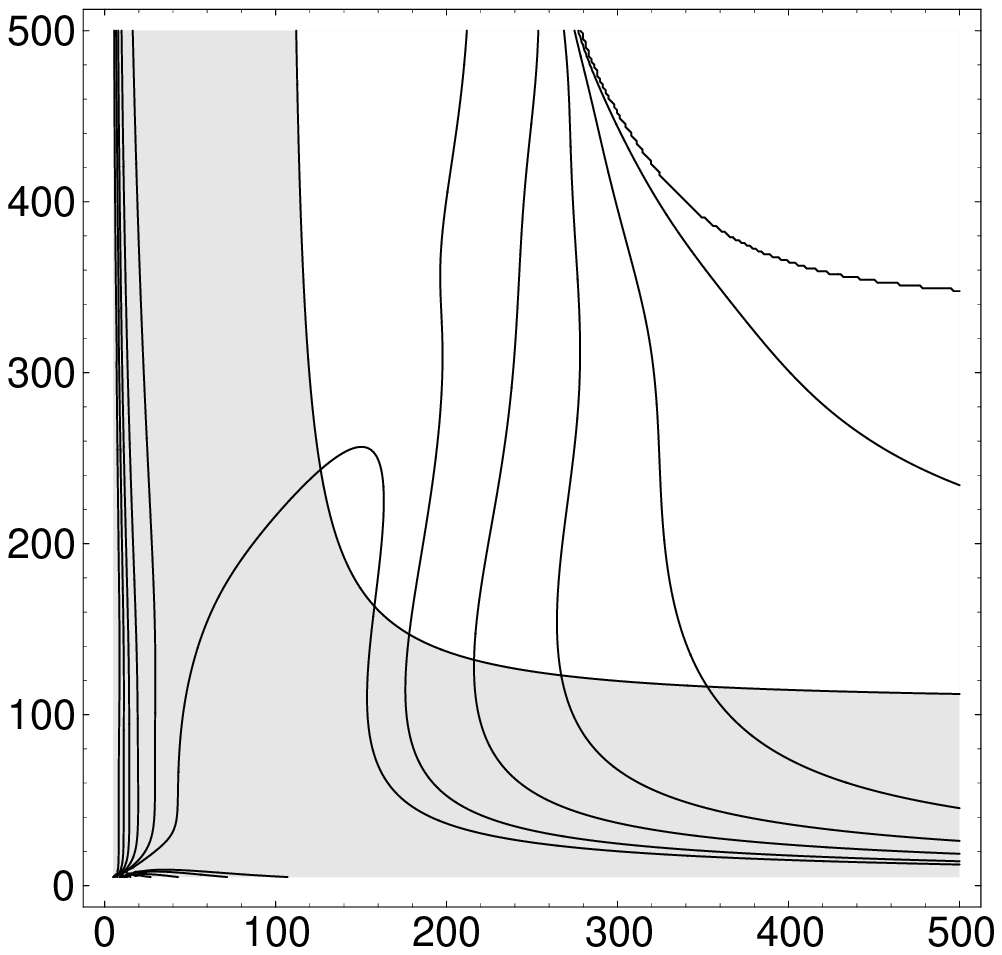}}
   \put(5.5,8.8){$|\mu|$~/GeV}
   \put(0,16.3){$M_2$~/GeV }
	\put(6,15){\begin{picture}(1,1)(0,0)
			\CArc(0,0)(6,0,380)
			\Text(0,0)[c]{{\scriptsize A}}
		\end{picture}}
\put(1.8,12.1){ \footnotesize$300$}
\put(2.6,13.5){\footnotesize$200$}
\put(3.2,13){\scriptsize$100$}
\put(3.7,12.5){\footnotesize$50$}
\put(4.3,12){\footnotesize$25$}
\put(5.9,12.5){\footnotesize$10$}
\put(0.5,8.8){Fig.~\ref{plots_12}a}
   \put(8,9){\includegraphics{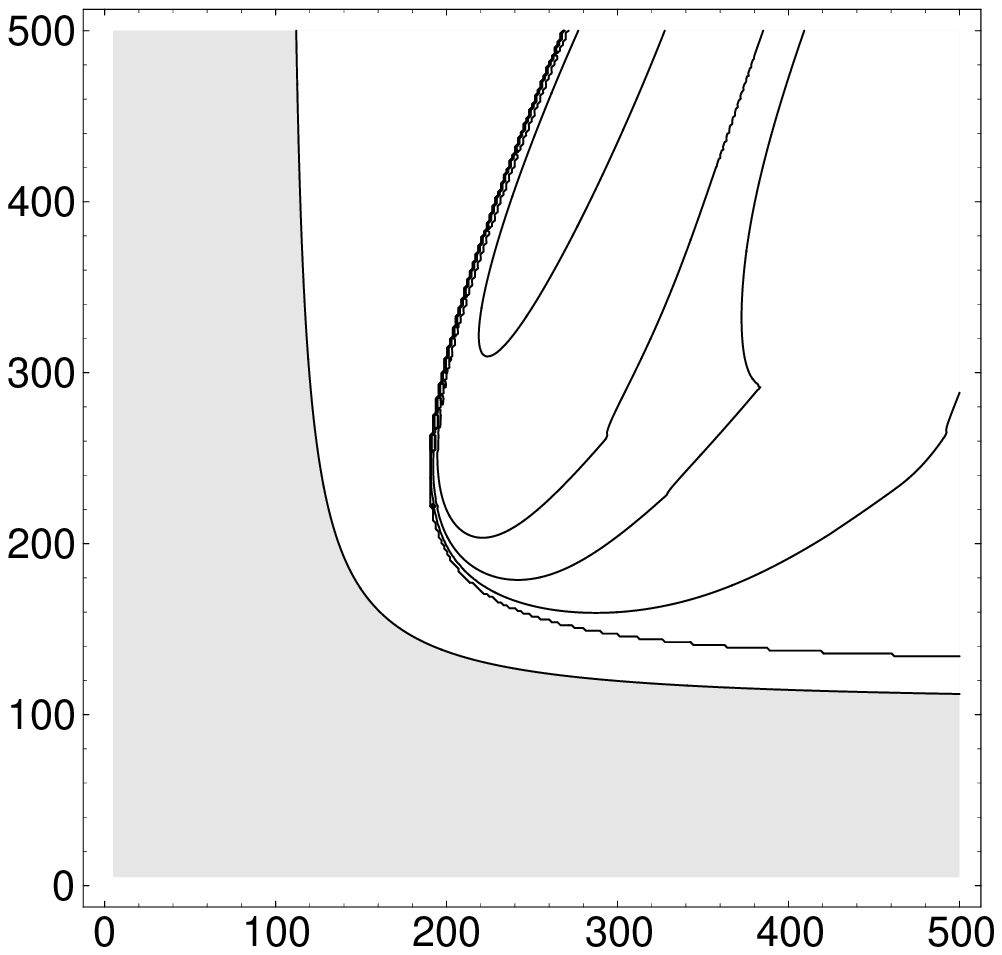}}
	\put(10.5,16.5){\fbox{BR$(\tilde{\chi}^0_2 \to \tilde{\ell}_R\ell_1)$ in \%}}
   \put(13.5,8.8){$|\mu|$~/GeV}
	\put(8,16.3){$M_2$~/GeV }
	\put(10.85,15){\begin{picture}(1,1)(0,0)
			\CArc(0,0)(6,0,380)
			\Text(0,0)[c]{{\scriptsize B}}
		\end{picture}}
\put(11.8,14.7){$64$}
\put(12.1,13.5){$40$}
\put(12.7,13){$20$}
\put(14,12.5){$4$}
\put(8.5,8.8){Fig.~\ref{plots_12}b}
	\put(0,0){\includegraphics{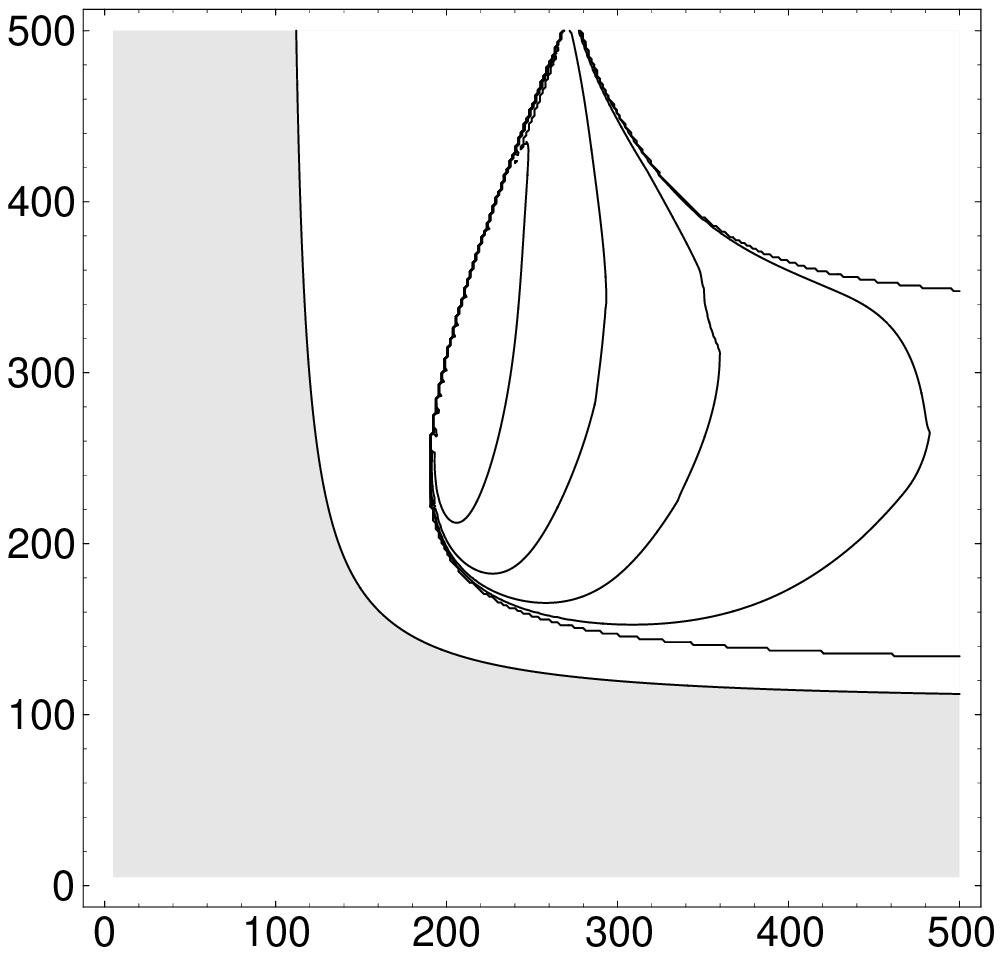}}
	\put(2.,7.5){\fbox{$\sigma(e^+e^- \to\tilde{\chi}^0_1
			\tilde{\chi}^0_1 \ell_1 \ell_2 )$ in fb}}
	\put(5.5,-0.3){$|\mu|$~/GeV}
	\put(0,7.3){$M_2$~/GeV }
		\put(6,6){\begin{picture}(1,1)(0,0)
			\CArc(0,0)(6,0,380)
			\Text(0,0)[c]{{\scriptsize A}}
	\end{picture}}
		\put(2.85,6){\begin{picture}(1,1)(0,0)
			\CArc(0,0)(6,0,380)
			\Text(0,0)[c]{{\scriptsize B}}
		\end{picture}}
	\put(3.2,4.2){\footnotesize $60$}
   \put(3.8,3.8){\footnotesize $20$}
   \put(4.6,3.5){\footnotesize $4 $}
	\put(5.8,3.4){\footnotesize $0.4 $}
	\put(0.5,-0.3){Fig.~\ref{plots_12}c}
   \put(8,0){\includegraphics{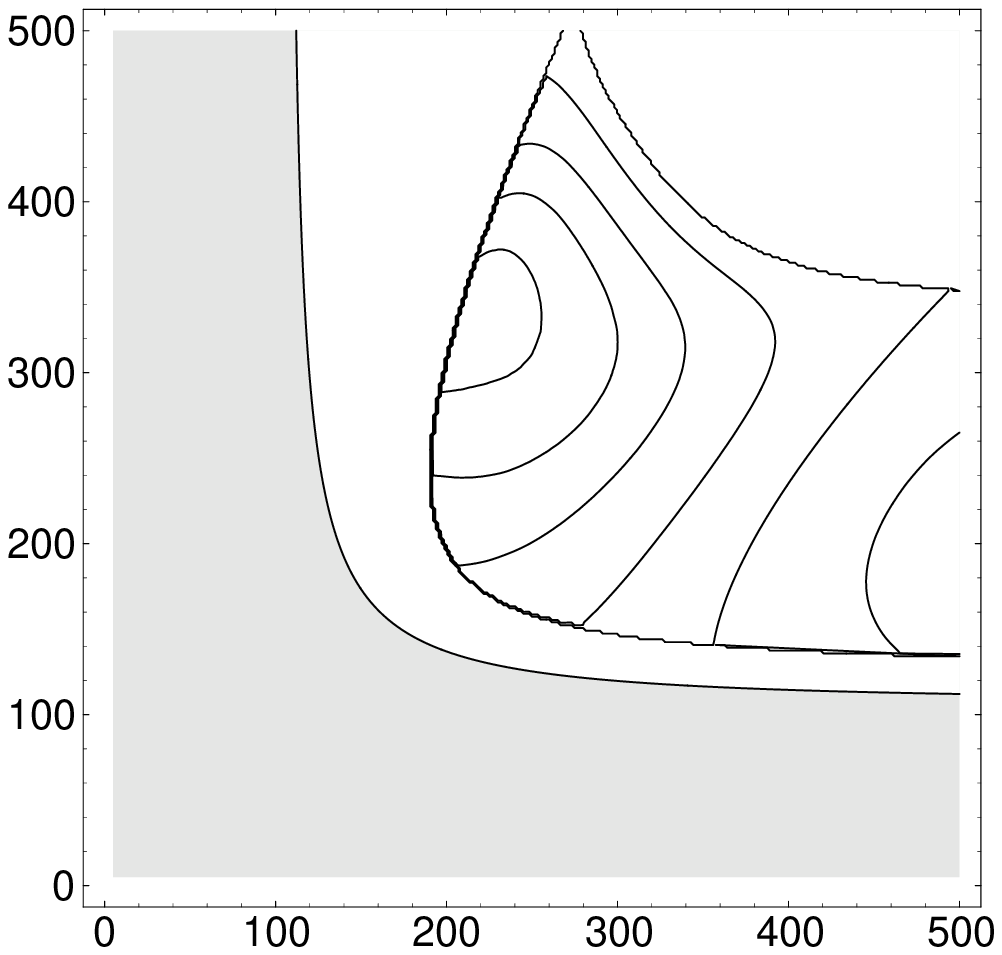}}
	\put(11.5,7.5){\fbox{${\mathcal A}_{II}$ in \% }}
	\put(13.5,-0.3){$|\mu|$~/GeV}
		\put(14,6){\begin{picture}(1,1)(0,0)
			\CArc(0,0)(6,0,380)
			\Text(0,0)[c]{{\scriptsize A}}
	\end{picture}}
		\put(10.85,6){\begin{picture}(1,1)(0,0)
			\CArc(0,0)(6,0,380)
			\Text(0,0)[c]{{\scriptsize B}}
		\end{picture}}
	\put(11.3,4.5){$ 9.5 $}
	\put(11.7,3.9){$ 8 $}
	\put(12.,3.5){$6 $}
	\put(12.4,3.2){$3 $}
   \put(13,3){$0 $}
   \put(14.3,2.8){$-3 $}
   \put(8.5,-0.3){Fig.~\ref{plots_12}d}
 \end{picture}
\vspace*{-1.5cm}
\caption{
	Contour plots for  
	\ref{plots_12}a: $\sigma(e^+e^- \to\tilde{\chi}^0_1\tilde{\chi}^0_2)$, 
	\ref{plots_12}b: BR$(\tilde{\chi}^0_2 \to \tilde{\ell}_R\ell_1)$,
	$ \ell= e,\mu$,
	\ref{plots_12}c: $\sigma(e^+e^-\to\tilde\chi^0_1\tilde\chi^0_2 ) \times
	{\rm BR}(\tilde \chi^0_2\to\tilde\ell_R\ell_1)\times
	{\rm BR}(\tilde\ell_R\to\tilde\chi^0_1\ell_2)$
	with BR($ \tilde\ell_R \to\tilde\chi^0_1\ell_2$) = 1,
	\ref{plots_12}d: the asymmetry ${\mathcal A}_{II}$,
	in the $|\mu|$--$M_2$ plane for $\varphi_{M_1}=0.5\pi $, 
	$\varphi_{\mu}=0$, taking  $\tan \beta=10$, $m_0=100$ GeV,
	$A_{\tau}=-250$ GeV, $\sqrt{s}=500$ GeV, $P_-=0.8$ and $P_+=-0.6$.
	The area A (B) is kinematically forbidden by
	$m_{\tilde\chi^0_1}+m_{\tilde\chi^0_2}>\sqrt{s}$
	$(m_{\tilde\ell_R}>m_{\tilde\chi^0_2})$.
	The gray  area is excluded by $m_{\tilde\chi_1^{\pm}}<104 $ GeV. 
	\label{plots_12}}
\end{figure}

In Fig.~\ref{plots_12}a we show the cross section for
$\tilde\chi^0_1 \tilde\chi^0_2$ production
for $\varphi_{\mu}=0$ and $\varphi_{M_1}=0.5~\pi$ in the $|\mu|$--$M_2$
plane. 
The cross section reaches values up to 300 fb.
For $|\mu|  \lsim 250 $ GeV the right selectron exchange dominates
so that our choice of polarization $P_-=0.8$ and $P_+=-0.6$
enhances the cross section by a factor as large as 2.5 compared 
to the unpolarized
case. For  $|\mu| \gsim 300$ GeV the left selectron exchange dominates 
because of the larger $\tilde\chi^0_2-\tilde e_L$ coupling.
In this region a sign reversal of both polarizations would enhance the
cross section by a factor between 1 and 20.

\begin{figure}[h]
	\begin{picture}(20,20)(0,-2)
	\put(2.5,7.5){\fbox{BR$(\tilde{\chi}^0_2 \to\tilde{\tau}_1\tau)$ in \%}}
	\put(0,0){\includegraphics{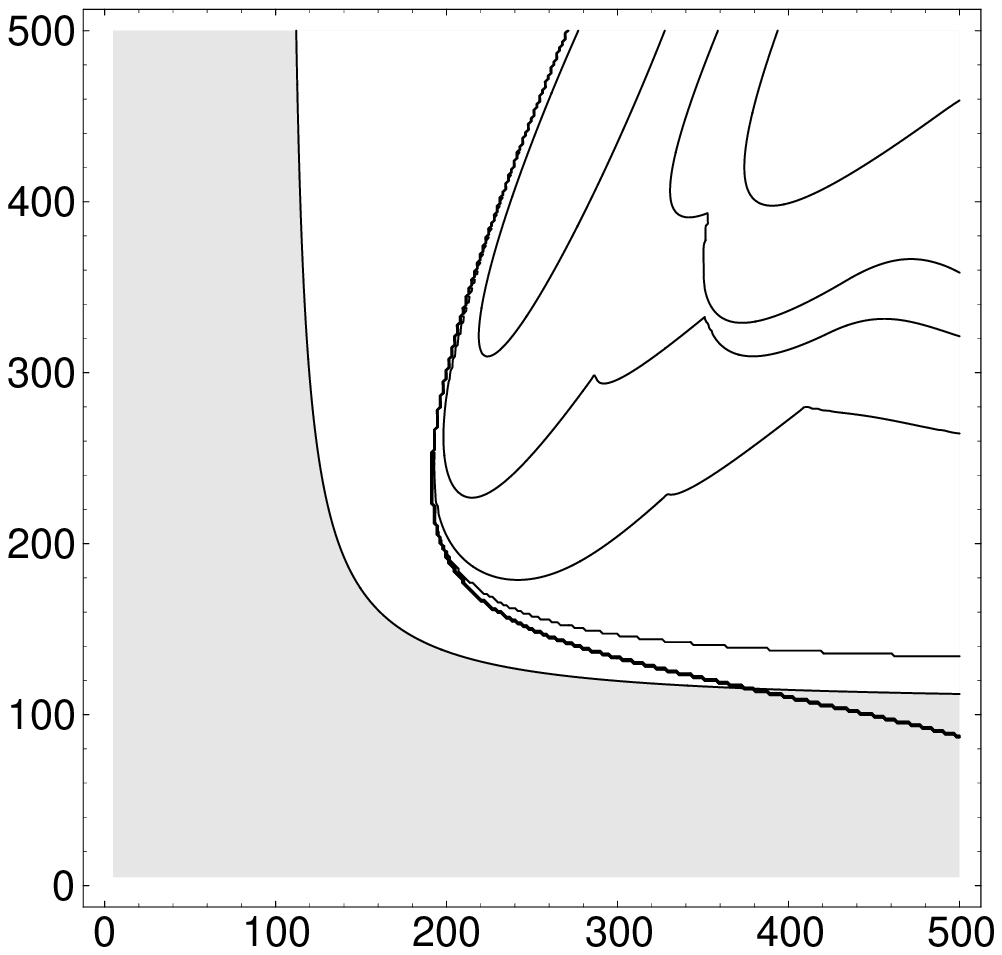}}
   \put(5.5,-0.3){$|\mu|$~/GeV}
   \put(0,7.3){$M_2$~/GeV }
\put(5.7,5.8){\footnotesize$15$}
\put(5.3,4.9){\footnotesize$40$}
\put(3.5,5.){\footnotesize$40$}
\put(4.5,4.1){\footnotesize$50$}
\put(5.5,3.6){\footnotesize$80$}
\put(6.,2.1){\footnotesize$100$}
		\put(2.85,6){\begin{picture}(1,1)(0,0)
			\CArc(0,0)(6,0,380)
			\Text(0,0)[c]{{\scriptsize B}}
			\end{picture}}
\put(0.5,-0.3){Fig.~\ref{plotsstau_12}a}
   \put(8,0){\includegraphics{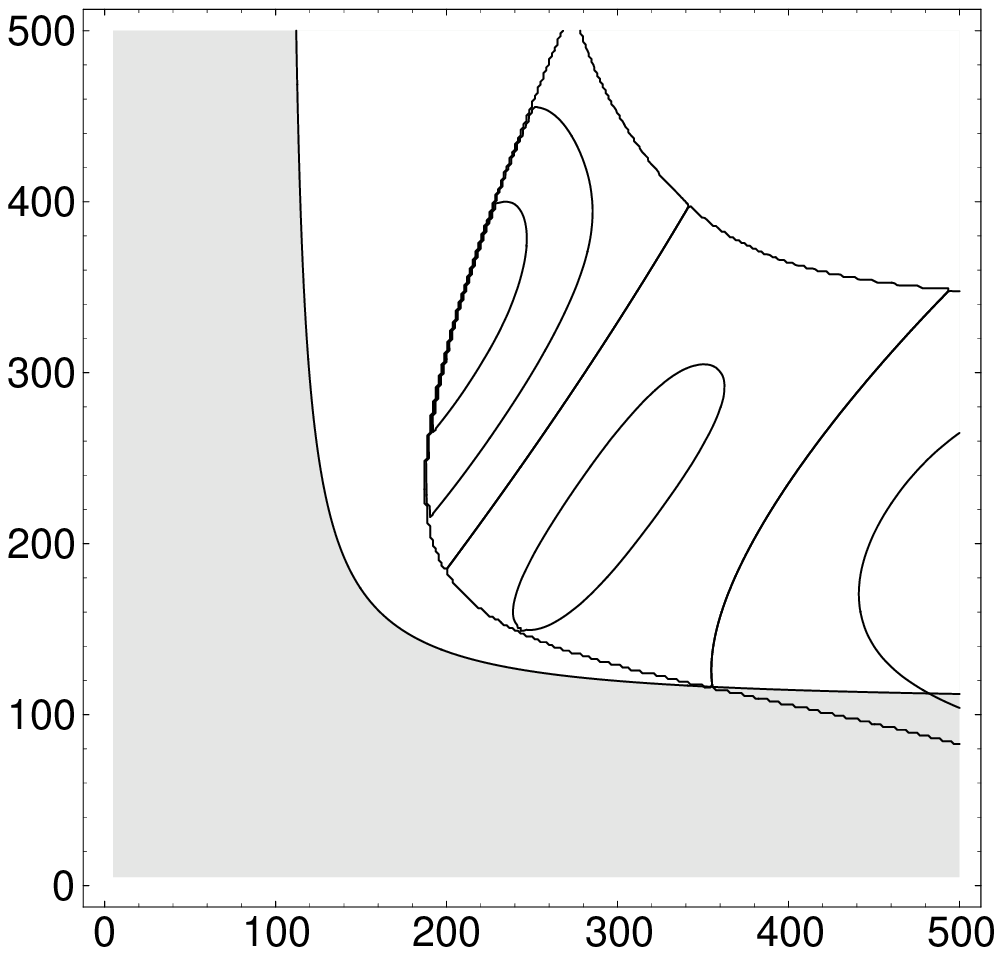}}
	\put(11.5,7.5){\fbox{${\mathcal A}_{II}$ in \% }}
	\put(13.5,-0.3){$|\mu|$~/GeV}
	\put(8,7.3){$M_2$~/GeV }
	\put(11.45,5.2){\scriptsize$ -6 $}
	\put(12.2,5.5){\scriptsize$ -3 $}
	\put(12.1,4){\footnotesize$0 $}
	\put(12.1,3.3){\footnotesize$3 $}
   \put(13,3){\footnotesize$0 $}
	\put(14.2,2.8){\footnotesize$-3 $}
		\put(13.5,-0.3){$|\mu|$~/GeV}
		\put(14,6){\begin{picture}(1,1)(0,0)
			\CArc(0,0)(6,0,380)
			\Text(0,0)[c]{{\scriptsize A}}
	\end{picture}}
		\put(10.85,6){\begin{picture}(1,1)(0,0)
			\CArc(0,0)(6,0,380)
			\Text(0,0)[c]{{\scriptsize B}}
			\end{picture}}
	\put(8.5,-0.3){Fig.~\ref{plotsstau_12}b}
 \end{picture}
\vspace*{-1.5cm}
\caption{
	Contour plots for  
	\ref{plotsstau_12}a: BR$(\tilde{\chi}^0_2 \to\tilde{\tau}_1\tau)$ and
	\ref{plotsstau_12}b: the asymmetry  ${\mathcal A}_{II}$,
	in the $|\mu|$--$M_2$ plane for $\varphi_{M_1}=0.5\pi $, 
	$\varphi_{\mu}=0$ and $A_{\tau}=-250$ GeV, taking  $\tan \beta=10$, 
	$m_0=100$ GeV, $\sqrt{s}=500$ GeV, $P_-=0.8$ and $P_-=-0.6$.
	The area A (B) is kinematically forbidden by
	$m_{\tilde\chi^0_1}+m_{\tilde\chi^0_2}>\sqrt{s}$
	$(m_{\tilde\tau_1}>m_{\tilde\chi^0_2})$.
	The gray  area is excluded by $m_{\tilde\chi_1^{\pm}}<104 $ GeV.
\label{plotsstau_12}}
\end{figure}

The branching ratio for the neutralino two-body decay into right
selectrons and smuons 
${\rm BR} (\tilde\chi^0_2 \to\tilde\ell_R\ell_1 )$
(summed over both signs of charge) is shown
in Fig.~\ref{plots_12}b. 
The branching ratio reaches values up to 64\% and decreases
with increasing $|\mu|$ when the two-body decays
into the lightest neutral Higgs boson $h^0$  and/or the
$Z$ boson are kinematically allowed. 
The channels into the $W$ boson do not open.
With our choice for the slepton masses, Eqs.~(\ref{RGE1}) and
(\ref{RGE2}), the decay into left selectrons and smuons can be neglected
because these channels are either not open or the branching fraction is
smaller than 1\%. As we assume that the squarks and the other
Higgs bosons are heavy, the decay into the stau is a competing channel,
which is discussed below. In our scenario this decay mode dominates 
for $M_2 \lsim 200$ GeV, see Fig.~\ref{plotsstau_12}a.
The resulting cross section 
$\sigma(e^+e^-\to\tilde\chi^0_1\tilde\chi^0_2 ) \times
{\rm BR}(\tilde \chi^0_2\to\tilde\ell_R\ell_1)\times
{\rm BR}(\tilde\ell_R\to\tilde\chi^0_1\ell_2)$
with BR($ \tilde\ell_R \to\tilde\chi^0_1\ell_2$) = 1
is shown in Fig.~\ref{plots_12}c.

Fig.~\ref{plots_12}d shows the $|\mu|$--$M_2$ dependence 
of the asymmetry ${\mathcal A}_{II}$ for $\varphi_{M_1}=0.5~\pi $ and $\varphi_{\mu}=0$.
In the region $|\mu|  \lsim 250 $ GeV, where the right selectron
exchange dominates, the asymmetry reaches $9.5\%$ for our choice of beam 
polarization. This enhances the asymmetry up to a factor of 2 compared to the
case of unpolarized beams. With increasing  $|\mu|$ the asymmetry
decreases and finally changes sign. This is due to the increasing
contributions of the left selectron exchange which contributes to the 
asymmetry with opposite sign and dominates for $|\mu| \gsim 300$ GeV.
In this region the asymmetry could be enhanced up to a factor 2 
by reversing the sign of both beam polarizations.

\begin{figure}[h]
 \begin{picture}(20,20)(0,-2)
	 \put(2.,16.5){\fbox{$\sigma(e^+\,e^- \to\tilde{\chi}^0_1 \,
			\tilde{\chi}^0_1 \ell_1 \,\ell_2 )$ in fb}}
	\put(0,9){\includegraphics{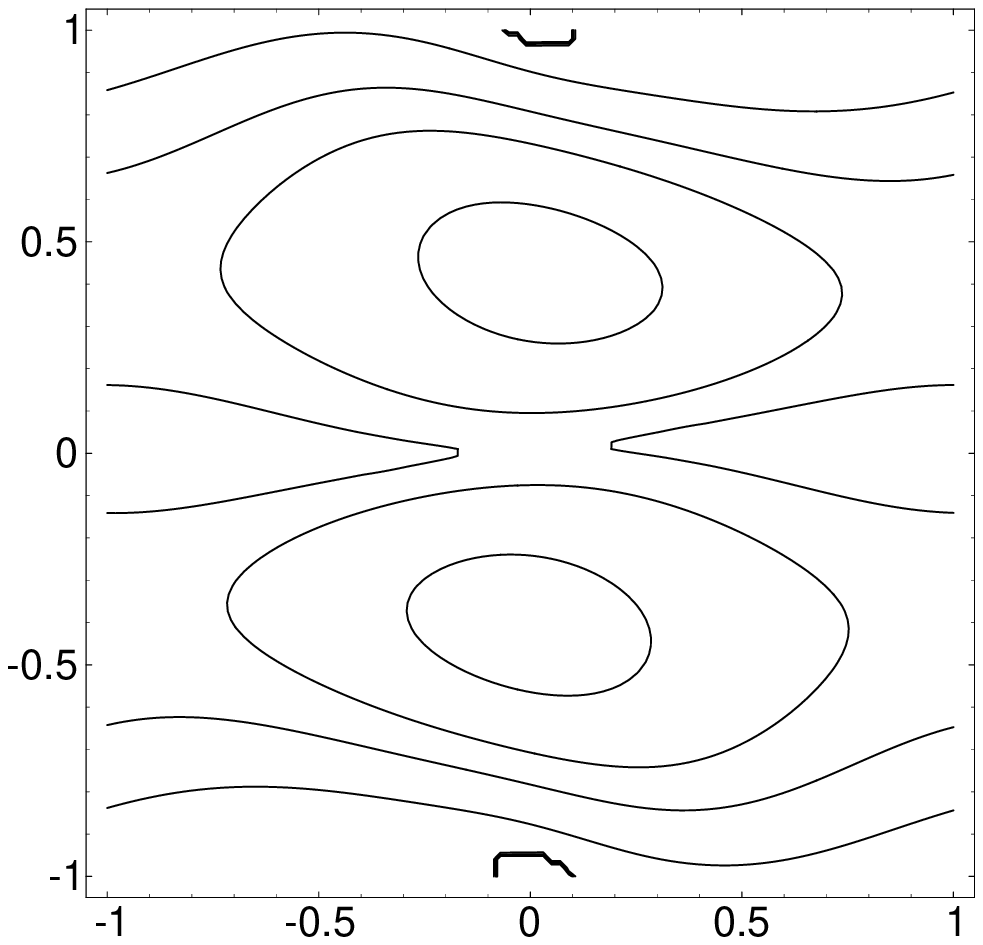}}
\put(6.5,8.8){$\varphi_{\mu}/\pi$}
\put(0,16.3){$ \varphi_{M_1}/\pi$ }
\put(4.,14.1){\footnotesize$64$}
\put(5.05,14.15){\footnotesize$56$}
\put(6.,14.4){\footnotesize$48$}
\put(6.6,14.95){\footnotesize$32$}
\put(6.3,12.9){\footnotesize$48$}
	\put(4.,11.2){\footnotesize$64$}
	\put(5.05,10.8){\footnotesize$56$}
	\put(5.8,10.55){\footnotesize$48$}
	\put(6.6,10.25){\footnotesize$32$}
	\put(1.3,12.45){\footnotesize$48$}
\put(0.5,8.8){Fig.~\ref{varphases_12}a}
\put(8,9){\includegraphics{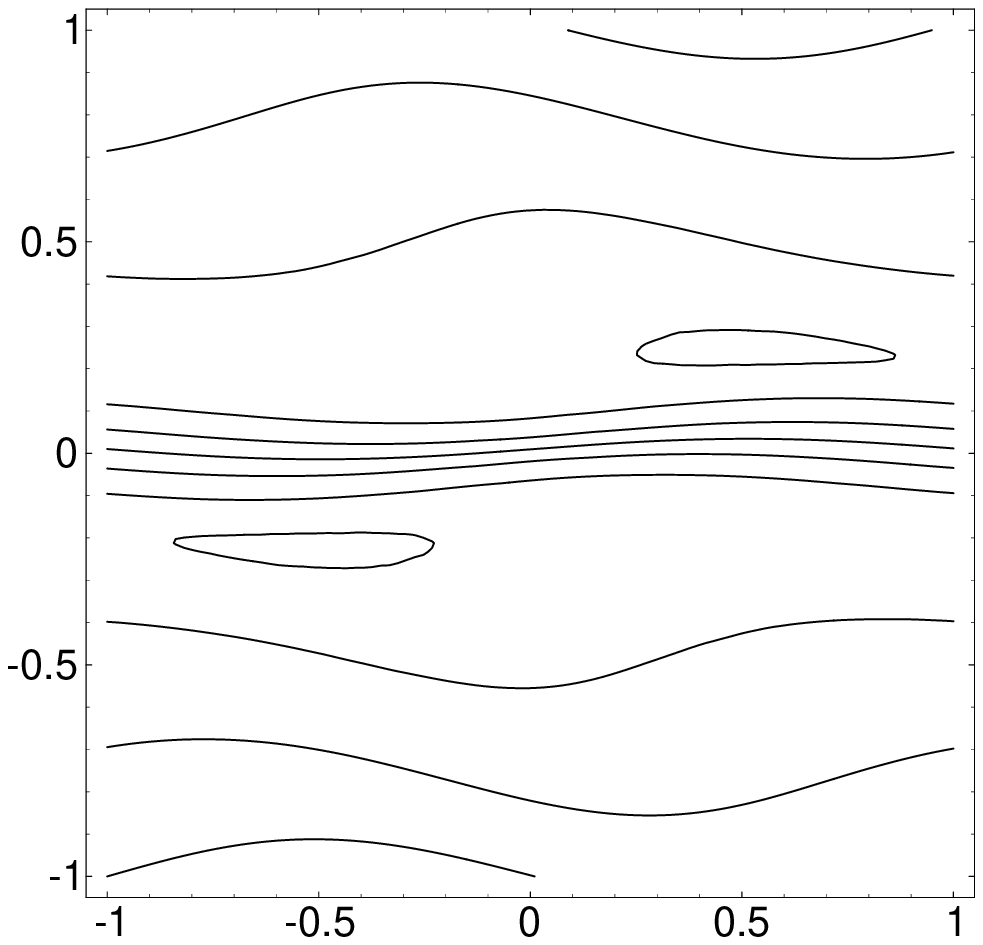}}
	\put(11.5,16.5){\fbox{${\mathcal A}_{II}$ in \% }}
   \put(14.5,8.8){$\varphi_{\mu}/\pi$}
   \put(8,16.3){$ \varphi_{M_1}/\pi$ }
	\put(13.5,15.6){\footnotesize$0$}
   \put(11.,15.1){\footnotesize$5$}
   \put(12.,14.1){\footnotesize$8$}
	\put(13.3,13.4){\footnotesize$8.9$}
	\put(12.,13.1){\footnotesize$8$}
	\put(10.,12.){\footnotesize$-8.9$}
	\put(12.,12.3){\footnotesize$-8$}
	   \put(12.,11.3){\footnotesize$-8$}
		\put(13,10.4){\footnotesize$-5$}
		\put(10.2,9.7){\footnotesize$0$}
		\put(8.5,8.8){Fig.~\ref{varphases_12}b}
	\put(8,0){\includegraphics{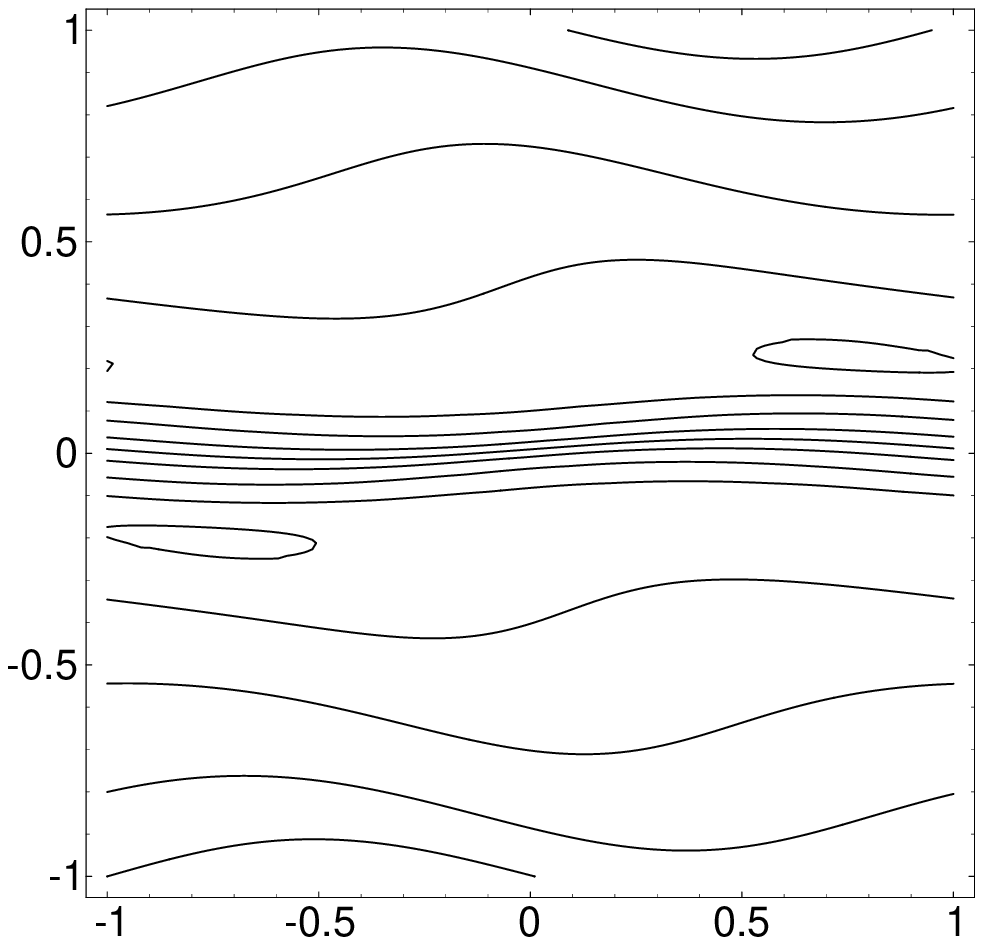}}
	\put(11.6,7.5){\fbox{${\mathcal A}_{I}$ in \% }}
	\put(14.5,-.3){$\varphi_{\mu}/\pi$}
	\put(8,7.3){$ \varphi_{M_1}/\pi$ }
	\put(11.2,6.25){\footnotesize$10$}
\put(11.5,5.6){\footnotesize$20$}
\put(12,4.8){\footnotesize$25$}
\put(14,4.35){\scriptsize$27$}
\put(12.6,4.2){\footnotesize$25$}
\put(13.5,6.6){\footnotesize$0$}
	\put(9.3,3.05){\scriptsize$-27$}
	\put(11.3,3.2){\footnotesize$-25$}
	\put(11.3,2.6){\footnotesize$-25$}
	\put(12.3,1.8){\footnotesize$-20$}
	\put(13.3,1.2){\footnotesize$-10$}
	\put(10.3,0.7){\footnotesize$0$}
	\put(0.5,-0.3){Fig.~\ref{varphases_12}c}
		\put(0,0){\includegraphics{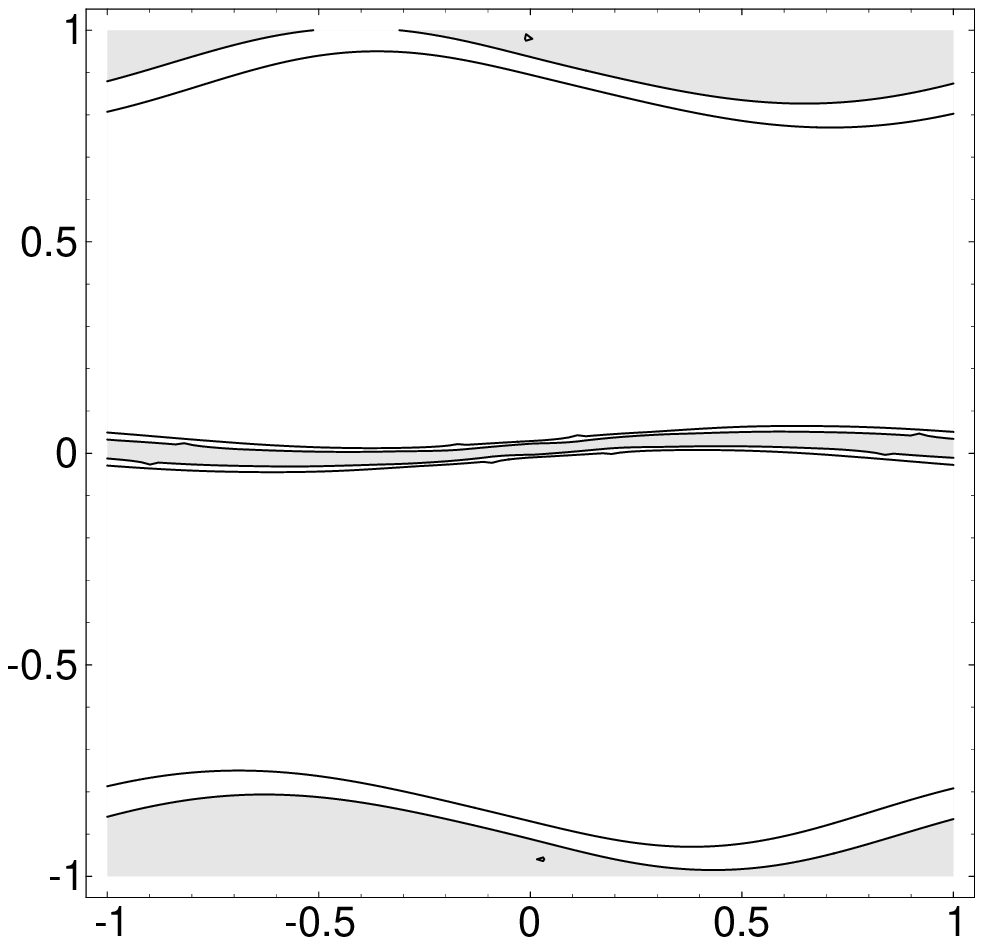}}
	\put(3.,7.5){\fbox{ $S={\mathcal A}_{II}\sqrt{N}$  }}
	\put(6.5,-.3){$\varphi_{\mu}/\pi$}
	\put(0,7.3){$ \varphi_{M_1}/\pi$ }
	\put(5.6,6.3){\footnotesize$3$}
	\put(5,5.75){\footnotesize$5$}
\put(3.8,3.3){\footnotesize$5$}
\put(4.2,4.1){\footnotesize$5$}
	\put(2.8,1.55){\footnotesize$5$}
	\put(2.3,0.95){\footnotesize$3$}
	\put(8.5,-0.3){Fig.~\ref{varphases_12}d}
 \end{picture}
\vspace*{-1.5cm}
\caption{
	Contour plots for  
	\ref{varphases_12}a: $\sigma(e^+e^-\to\tilde\chi^0_1\tilde\chi^0_2 ) \times
	{\rm BR}(\tilde \chi^0_2\to\tilde\ell_R\ell_1)\times
	{\rm BR}(\tilde\ell_R\to\tilde\chi^0_1\ell_2)$
	with BR($ \tilde\ell_R \to\tilde\chi^0_1\ell_2$) = 1,
	\ref{varphases_12}b: the asymmetry ${\mathcal A}_{II}$,
	\ref{varphases_12}c: the standard deviation $S$,
	\ref{varphases_12}d: the asymmetry ${\mathcal A}_{I}$,
	in the $\varphi_{\mu}$--$\varphi_{M_1}$ plane 
	for  $M_2=400$ GeV and $|\mu|=240$ GeV,
	taking  $\tan \beta=10$, $m_0=100$ GeV,
	$A_{\tau}=-250$ GeV, $\sqrt{s}=500$ GeV, $P_-=0.8$ and $P_+=-0.6$.
	For $\varphi_{M_1},\varphi_{\mu}=0$ we have
	$m_{\tilde \ell_R}=221$ GeV, $m_{\tilde\chi_1^0}=178$ GeV and
	$m_{\tilde\chi_2^0}=243$ GeV.
\label{varphases_12}}
\end{figure}

The sensitivity of the cross section $\sigma$ and the asymmetry 
${\mathcal A}_{II}$ on the CP phases can be seen  by contour plots in the
$\varphi_{\mu}$--$\varphi_{M_1}$ plane,  for $|\mu|=240$ GeV
and  $M_2=400$ GeV (Fig.~\ref{varphases_12}).
In our scenario the variation of the cross section 
(Fig.~\ref{varphases_12}a) is more than 100$\%$. 
In addition to the CP sensitive observables, the cross section may 
also serve to determine the values of the phases.
Using unpolarized beams, the cross section would be reduced
by a factor 0.4.
The asymmetry ${\mathcal A}_{II}$ (Fig.~\ref{varphases_12}b)
varies between  -8.9$\%$ and 8.9$\%$. It is remarkable
that these maximal values are not necessarily obtained 
for maximal  CP phases. In our scenario the asymmetry is 
much more sensitive to variations of the phase $\varphi_{M_1}$ 
around  $0$. 
On the other hand, the asymmetry is rather insensitive to $\varphi_{\mu}$. 
For unpolarized beams this asymmetry would be reduced roughly by a factor 0.33.

The relative statistical error of each asymmetry 
${\mathcal A}$ can be calculated to $\delta {\mathcal A} = 
\Delta {\mathcal A}/{\mathcal A} = S/({\mathcal A} \sqrt{N})$,
with $S$ standard deviations,
assuming a Gaussian distribution of the asymmetry ${\mathcal A}$.
Here, $N={\mathcal L} \sigma$ is the number of events with 
${\mathcal L}$ the total integrated luminosity and 
$\sigma$ the total cross section. Assuming $\delta {\mathcal A}
\approx1$, it follows $S \approx {\mathcal A} \sqrt{N}$.
For example, in order to measure an asymmetry of 5\% with S=2
(confidence level of 95\%), one would need at least $1.5\times 10^3$ events.
This corresponds to a total cross section for reactions
(\ref{production})-(\ref{decay_2}) of 3.1 fb with 
${\mathcal L}=500~{\rm fb}^{-1}$. We show the contour lines of  
$S=3$ and 5 for ${\mathcal A}_{II}$ in Fig.~\ref{varphases_12}c with  
${\mathcal L}= 500$ fb$^{-1}$. In the gray shaded area $S<3$.

In Fig.~\ref{varphases_12}d we also show the asymmetry 
${\mathcal A}_{I}$ which is a factor 2.9  larger than 
${\mathcal A}_{II}$, because in ${\mathcal A}_{II}$ the CP-violating 
effect from the production is washed out by the kinematics of the 
slepton decay. However, for a measurement of ${\mathcal A}_{I}$ the 
reconstruction of the $\tilde{\chi}^0_2$ momentum is necessary.
The asymmetry ${\mathcal A}_{I}$ shows a similar dependence on the phases
as  ${\mathcal A}_{II}$ because both are due to the non vanishing neutralino 
polarization perpendicular to the production plane.

It is interesting to note that, due to the weak dependence on
$\varphi_{\mu}$,
the asymmetries can be sizable for $\varphi_{\mu} \approx 0$.
Small values for $\varphi_{\mu}$ are suggested by constraints on
electron and neutron electric dipole moments (EDMs) \cite{edmsexp} 
for a typical SUSY scale of the order of a few 100 GeV
(for a review see, e.g., \cite{edmstheo}).

Next we want to comment on the neutralino decay into the scalar tau and
discuss the main differences from the decay into the selectron and smuon.
In some regions of the parameter space, the decay
of the neutralino into the lightest stau $\tilde\tau_1$ may dominate over
that into the right selectron and smuon, and may even be the 
only decay channel. In Fig.~\ref{plotsstau_12}a we show the branching 
ratio $BR (\tilde\chi^0_2 \to \tilde\tau_1\tau)$ in the
$|\mu|$--$M_2 $ plane for $A_{\tau} = -250$ GeV,
$\varphi_{M_1}=0.5 \, \pi $ and $\varphi_{\mu}=0$.
For $M_2<200$ GeV the branching ratio 
$BR (\tilde\chi^0_2 \to \tilde\tau_1\tau)$ is larger than 80\%. 
However, due to the mixing in the stau sector the asymmetry 
${\mathcal A}_{II}$, Fig.~\ref{plotsstau_12}b, is reduced compared to that
in the selectron and smuon channels, see Fig.~\ref{plots_12}d. 
The reason is the suppression factor 
$(|a^{\tilde \tau}_{ki}|^2-|b^{\tilde \tau}_{ki}|^2)/(
  |a^{\tilde \tau}_{ki}|^2+|b^{\tilde \tau}_{ki}|^2)$, 
Eq.~(\ref{Amixing}), which may be small or even be zero. This may 
lead to a reduced or vanishing asymmetry, respectively,
even in the case of non zero CP phases. 

\subsection{Production of $\tilde\chi^0_1 \tilde\chi^0_3$ }

\begin{figure}[h]
 \begin{picture}(20,20)(0,-2)
	\put(0,0){\includegraphics{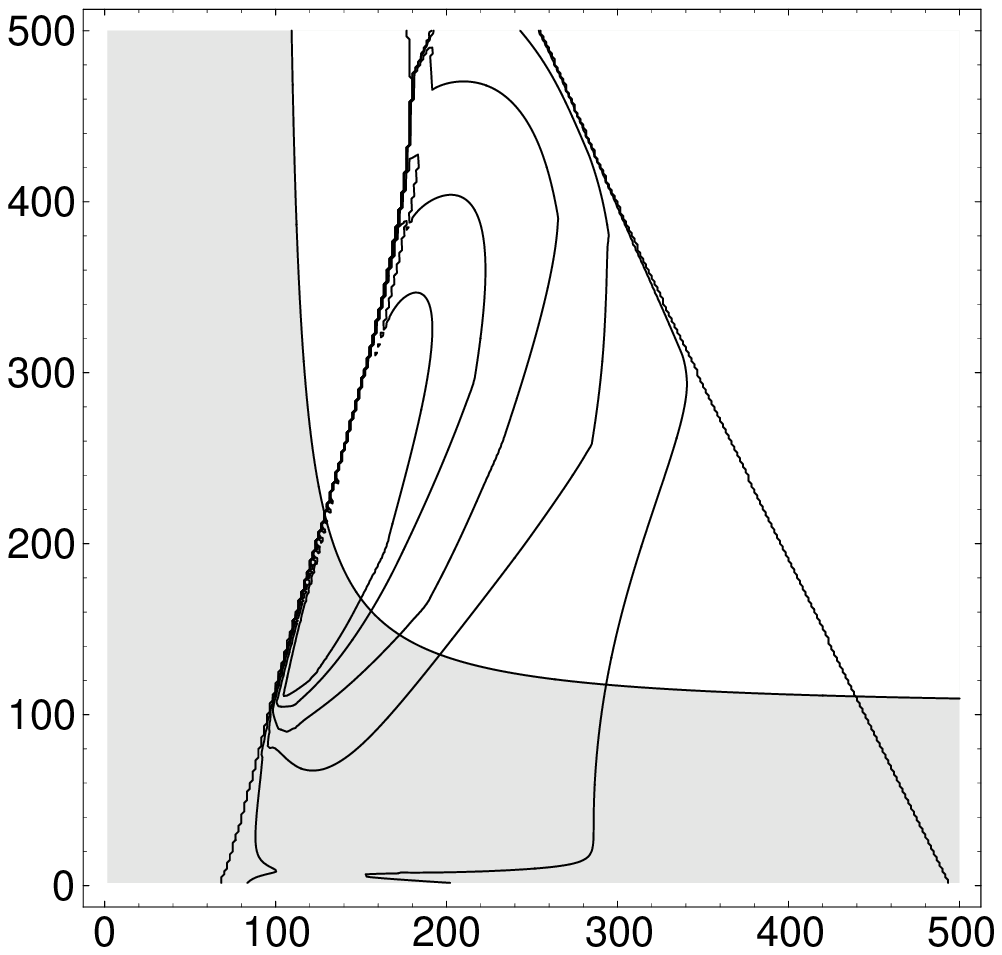}}
	\put(2.,7.5){\fbox{$\sigma(e^+\,e^- \to\tilde{\chi}^0_1 \,
			\tilde{\chi}^0_1 \ell_1 \,\ell_2 )$ in fb}}
	\put(5.5,-0.3){$|\mu|$~/GeV}
	\put(0,7.3){$M_2$~/GeV }
	\put(2.4,3.6){\scriptsize$ 100 $}
	\put(3.,5.2){\footnotesize$60 $}
	\put(3.5,5.7){\footnotesize$20 $}
	\put(3.9,3.7){\footnotesize$4 $}
	\put(4.,3.){\footnotesize$0.4 $}
			\put(6,6){\begin{picture}(1,1)(0,0)
			\CArc(0,0)(6,0,380)
			\Text(0,0)[c]{{\scriptsize A}}
	\end{picture}}
		\put(2.5,6){\begin{picture}(1,1)(0,0)
			\CArc(0,0)(6,0,380)
			\Text(0,0)[c]{{\scriptsize B}}
			\end{picture}}
	\put(0.5,-0.3){Fig.~\ref{plots_13}a}
   \put(8,0){\includegraphics{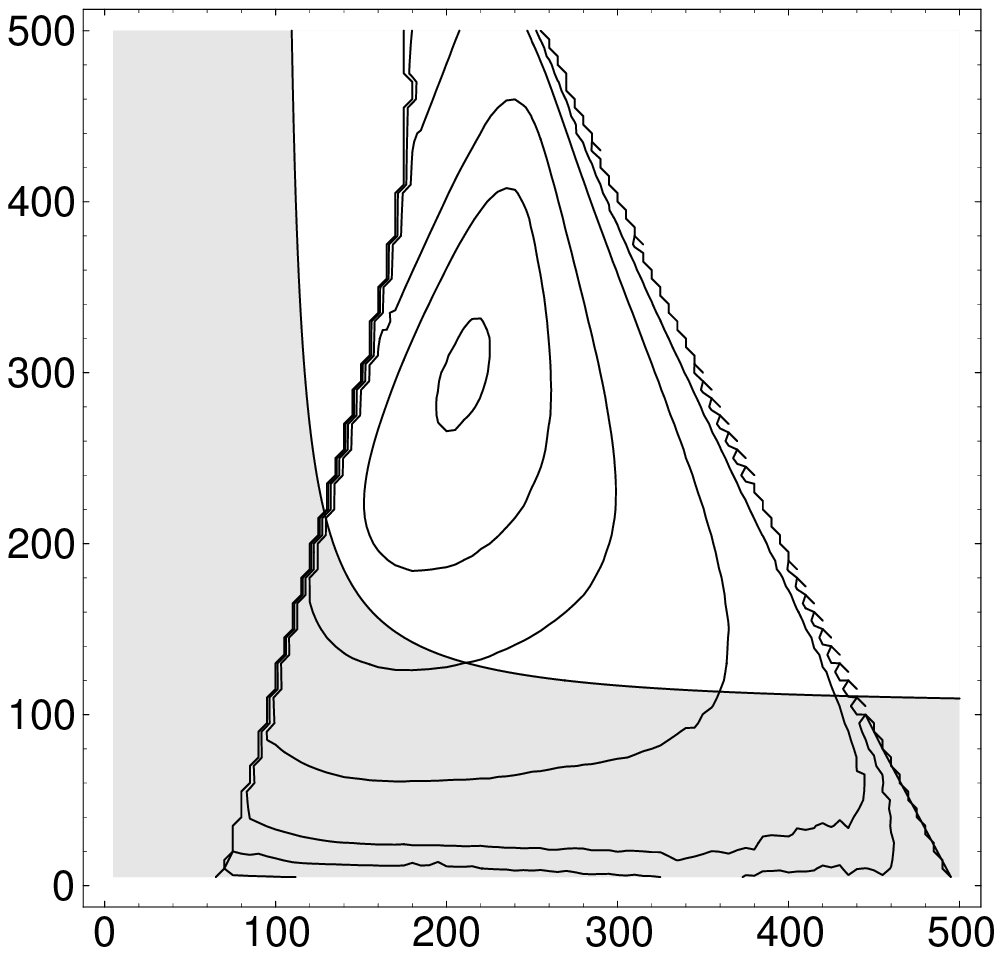}}
	\put(11.5,7.5){\fbox{${\mathcal A}_{II}$ in \% }}
	\put(13.5,-0.3){$|\mu|$~/GeV}
	\put(8,7.3){$M_2$~/GeV }
	\put(11.,4.3){\scriptsize$ -9.5 $}
	\put(11.2,3.4){\footnotesize$-8 $}
	\put(11.7,2.9){\footnotesize$-6 $}
   \put(12.6,2.3){\footnotesize$-3 $}
	\put(13.4,1.3){\footnotesize$-1 $}
	\put(14.4,0.8){\scriptsize$0 $}
			\put(14,6){\begin{picture}(1,1)(0,0)
			\CArc(0,0)(6,0,380)
			\Text(0,0)[c]{{\scriptsize A}}
	\end{picture}}
		\put(10.5,6){\begin{picture}(1,1)(0,0)
			\CArc(0,0)(6,0,380)
			\Text(0,0)[c]{{\scriptsize B}}
			\end{picture}}
	\put(8.5,-0.3){Fig.~\ref{plots_13}b}
 \end{picture}
\vspace*{-1.5cm}
\caption{
	Contour plots for  
	\ref{plots_13}a: $\sigma(e^+e^-\to\tilde\chi^0_1\tilde\chi^0_3) \times
	{\rm BR}(\tilde \chi^0_3\to\tilde\ell_R\ell_1)\times
	{\rm BR}(\tilde\ell_R\to\tilde\chi^0_1\ell_2)$
	with BR($ \tilde\ell_R \to\tilde\chi^0_1\ell_2$) = 1 and $\ell= e,\mu$,
	\ref{plots_13}b: the asymmetry ${\mathcal A}_{II}$,
	in the $|\mu|$--$M_2$ plane for $\varphi_{M_1}=0.5\pi $, 
	$\varphi_{\mu}=0$, taking  $\tan \beta=10$, $m_0=100$ GeV,
	$A_{\tau}=-250$ GeV, $\sqrt{s}=500$ GeV, $P_-=0.8$ and $P_+=-0.6$.
	The area A (B) is kinematically forbidden by
	$m_{\tilde\chi^0_1}+m_{\tilde\chi^0_3}>\sqrt{s}$
	$(m_{\tilde\ell_R}>m_{\tilde\chi^0_3})$.
	The gray  area is excluded by $m_{\tilde\chi_1^{\pm}}<104 $ GeV. 
\label{plots_13}}
\end{figure}

We show in Fig.~\ref{plots_13}a and b contour plots of the cross section 
$\sigma(e^+e^-\to\tilde\chi^0_1\tilde\chi^0_3 ) \times
{\rm BR}(\tilde\chi^0_3\to\tilde\ell_R\ell_1)\times
{\rm BR}(\tilde\ell_R\to\tilde\chi^0_1\ell_2)$
%{\bf
with BR($ \tilde\ell_R \to\tilde\chi^0_1\ell_2$) = 1
%}
and of the asymmetry ${\mathcal A}_{II}$, respectively.
The cross section with polarized beams reaches more than 100 fb,
which is up to a factor 2.5 larger than for unpolarized
beams. The asymmetry ${\mathcal A}_{II}$, shown in Fig.~\ref{plots_13}b, 
reaches -9.5\%. For unpolarized beams this value would be reduced 
by a factor 0.75.  For our choice of parameters the cross section and
the asymmetry for $\tilde\chi^0_1  \tilde\chi^0_3$ 
production and decay show a similar dependence on $M_2$ and $|\mu|$ 
as for $\tilde\chi^0_1 \tilde\chi^0_2$ production, however, the
kinematically allowed regions are different.
%{\bf
We also studied the $\varphi_{\mu}$ dependence of ${\mathcal A}_{II}$.
For $\varphi_{\mu}=0.5\pi (0.1\pi)$ and $\varphi_{M_1}=0$,
the maximal values of ${\mathcal A}_{II}$ in the
$M_2$--$|\mu|$ plane are $|{\mathcal A}_{II}|<3\%(1\%)$.
%}
\subsection{Production of $\tilde \chi ^0_2 \tilde \chi^0_3$ }

\begin{figure}[h]
 \begin{picture}(20,20)(0,-2)
	\put(2.5,16.5){\fbox{$\sigma(e^+\,e^- \to\tilde\chi^0_2 \,
			\tilde\chi^0_3)$ in fb}}
	\put(0,9){\includegraphics{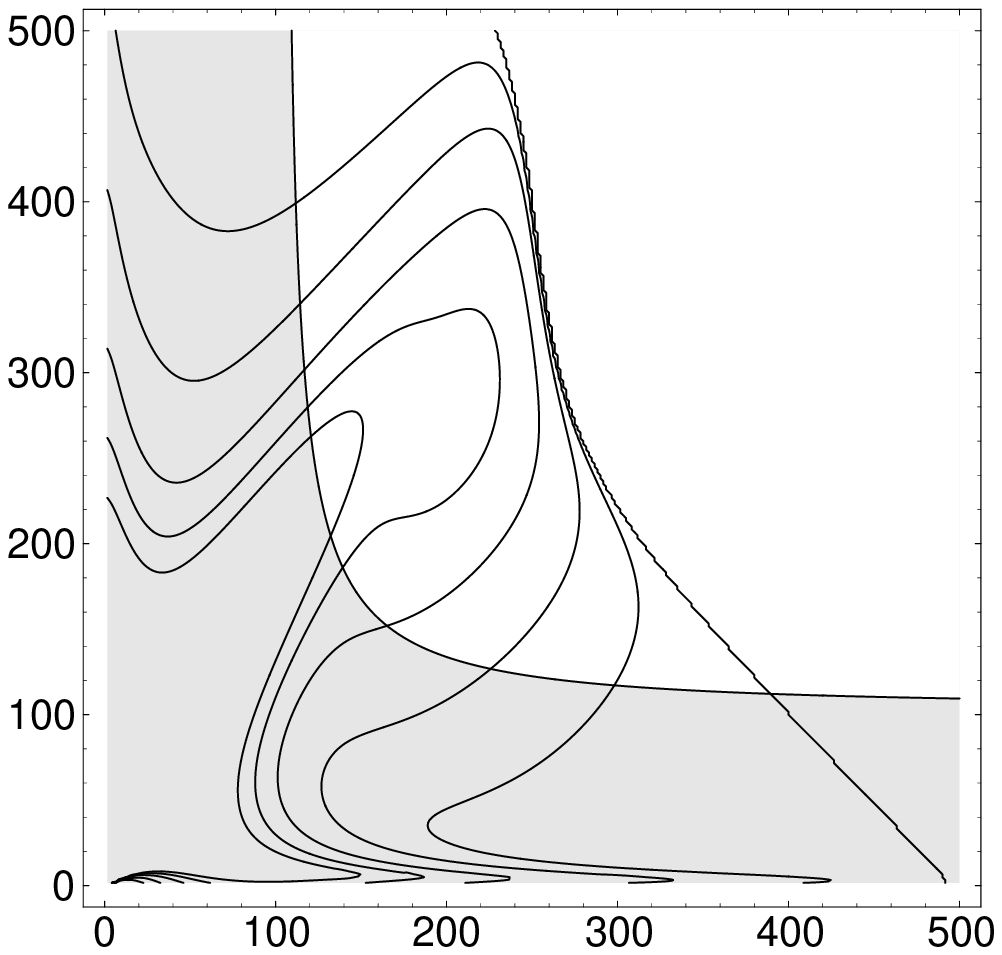}}
   \put(5.5,8.8){$|\mu|$~/GeV}
   \put(0,16.3){$M_2$~/GeV }
\put(1.8,12.3){\footnotesize$100$}
\put(2.9,12.65){\footnotesize$75$}
\put(3.1,12.1){\footnotesize$50$}
\put(3.5,11.7){\footnotesize$25$}
\put(4.1,11.3){\footnotesize$10$}
	\put(5.5,14){\begin{picture}(1,1)(0,0)
			\CArc(0,0)(6,0,380)
			\Text(0,0)[c]{{\scriptsize A}}
	\end{picture}}
\put(0.5,8.8){Fig.~\ref{plots_23}a}
   \put(8,9){\includegraphics{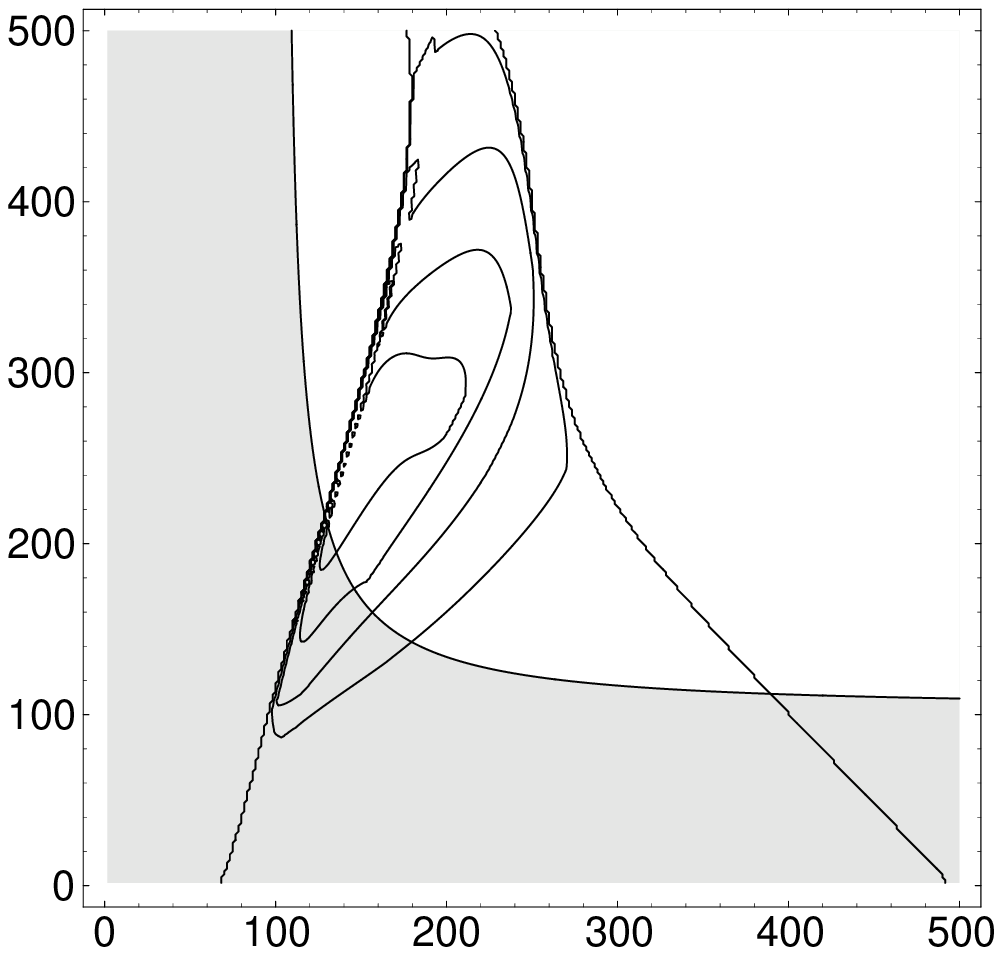}}
	\put(10.,16.5)
{\fbox{$\sigma(e^+\,e^- \to\tilde\chi^0_2 \,
			\tilde\chi^0_1 \ell_1 \,\ell_2 )$ in fb}}
   \put(13.5,8.8){$|\mu|$~/GeV}
   \put(8,16.3){$M_2$~/GeV }
\put(11.3,15.4){\footnotesize$4$}
\put(11.3,14.5){\footnotesize$20$}
\put(11.1,13.8){\footnotesize$40$}
\put(10.8,13.1){\footnotesize$52$}
	\put(13.5,14){\begin{picture}(1,1)(0,0)
			\CArc(0,0)(6,0,380)
			\Text(0,0)[c]{{\scriptsize A}}
	\end{picture}}
		\put(10.5,15){\begin{picture}(1,1)(0,0)
			\CArc(0,0)(6,0,380)
			\Text(0,0)[c]{{\scriptsize B}}
			\end{picture}}
\put(8.5,8.8){Fig.~\ref{plots_23}b}
	\put(0,0){\includegraphics{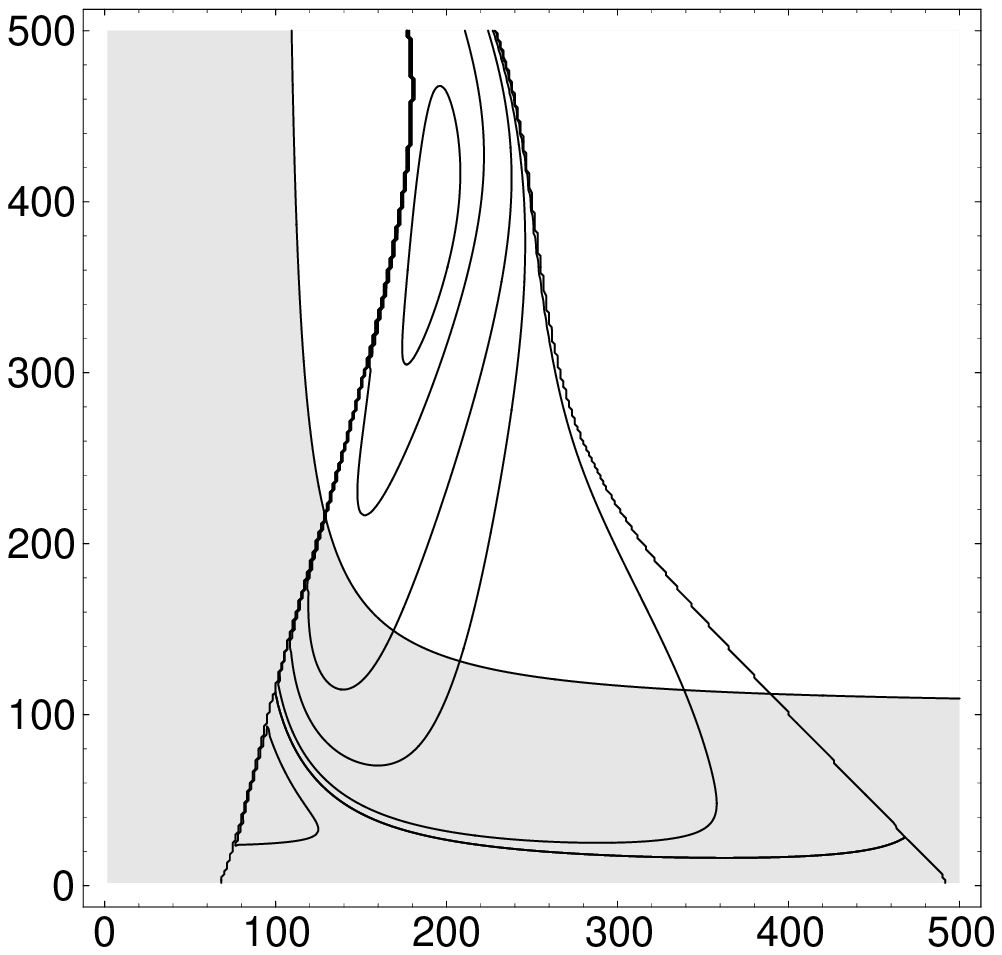}}
	\put(3.5,7.5){\fbox{${\mathcal A}_{I}$ in \% }}
	\put(5.5,-0.3){$|\mu|$~/GeV}
	\put(0,7.3){$M_2$~/GeV }
	\put(2.9,5.3){\footnotesize$ 25 $}
   \put(2.6,4){\footnotesize$15 $}
	\put(2.6,3.){\footnotesize$10 $}
	\put(3.2,2.7){\footnotesize$5 $}
	\put(4.5,2.2){\footnotesize$2 $}
	\put(5.5,1.05){\footnotesize$0 $}
	\put(1.7,1.15){\footnotesize$-3 $}
				\put(5.5,5){\begin{picture}(1,1)(0,0)
			\CArc(0,0)(6,0,380)
			\Text(0,0)[c]{{\scriptsize A}}
	\end{picture}}
		\put(2.5,6){\begin{picture}(1,1)(0,0)
			\CArc(0,0)(6,0,380)
			\Text(0,0)[c]{{\scriptsize B}}
			\end{picture}}
	\put(0.5,-0.3){Fig.~\ref{plots_23}c}
   \put(8,0){\includegraphics{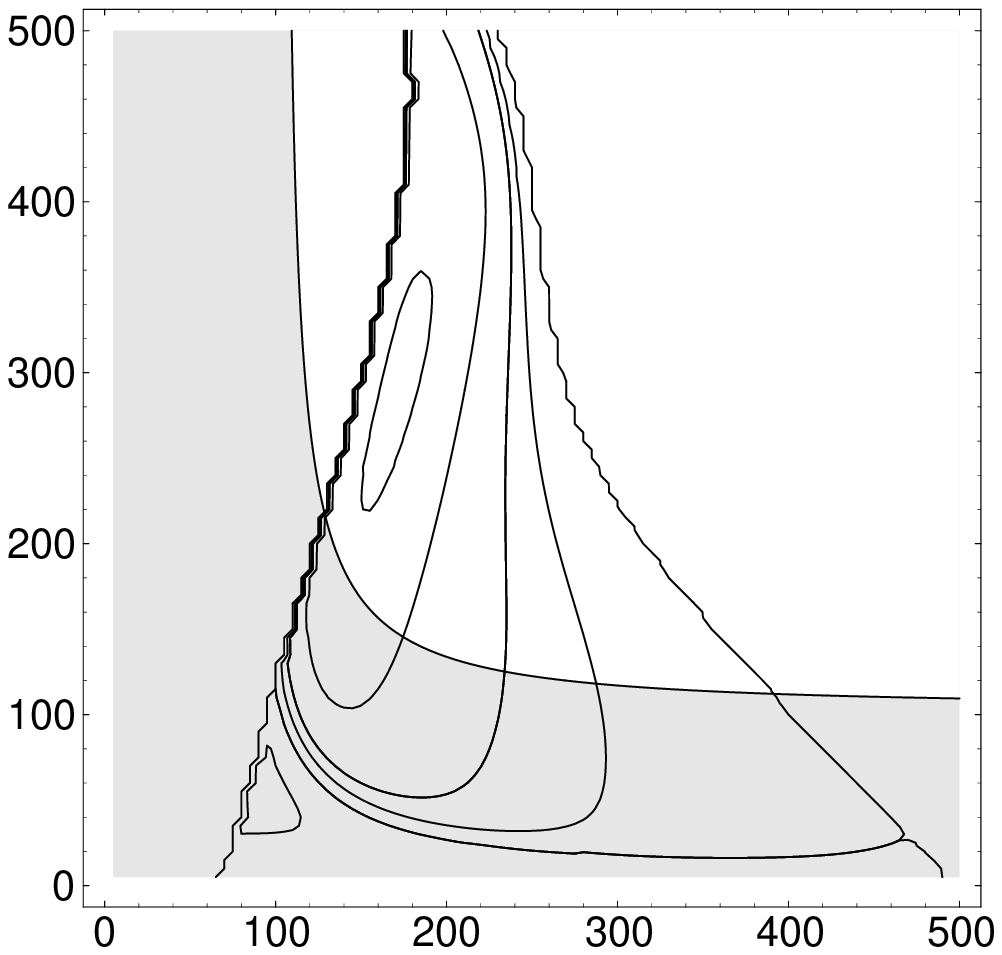}}
	\put(11.5,7.5){\fbox{${\mathcal A}_{II}$ in \% }}
	\put(13.5,-0.3){$|\mu|$~/GeV}
	\put(8,7.3){$M_2$~/GeV }
	\put(10.6,4.1){\footnotesize$ 10 $}
	\put(10.8,3){\footnotesize$5$}
	\put(11.4,2.55){\footnotesize$2 $}
	\put(11.9,2.3){\footnotesize$1 $}
	\put(13.5,1.1){\footnotesize$0 $}
	\put(9.6,1.2){\footnotesize$-2 $}
				\put(13.5,5){\begin{picture}(1,1)(0,0)
			\CArc(0,0)(6,0,380)
			\Text(0,0)[c]{{\scriptsize A}}
	\end{picture}}
		\put(10.5,6){\begin{picture}(1,1)(0,0)
			\CArc(0,0)(6,0,380)
			\Text(0,0)[c]{{\scriptsize B}}
			\end{picture}}
\put(8.5,-0.3){Fig.~\ref{plots_23}d}
 \end{picture}
\vspace*{-1.5cm}
\caption{
	Contour plots for  
	\ref{plots_23}a: $\sigma(e^+e^- \to\tilde{\chi}^0_2\tilde{\chi}^0_3)$, 
	\ref{plots_23}b: $\sigma(e^+e^-\to\tilde\chi^0_2\tilde\chi^0_3 ) \times
	{\rm BR}(\tilde \chi^0_3\to\tilde\ell_R\ell_1)\times
	{\rm BR}(\tilde\ell_R\to\tilde\chi^0_1\ell_2)$ for $ \ell= e,\mu$,
	and BR($\tilde\ell_R\to\tilde\chi^0_1\ell_2$) = 1,
	\ref{plots_23}c: the asymmetry ${\mathcal A}_{I}$,
	\ref{plots_23}d: the asymmetry ${\mathcal A}_{II}$,
	in the $|\mu|$--$M_2$ plane for $\varphi_{M_1}=0.5\pi $, 
	$\varphi_{\mu}=0$, taking  $\tan \beta=10$, $m_0=100$ GeV,
	$A_{\tau}=-250$ GeV, $\sqrt{s}=500$ GeV, $P_-=0.8$ and $P_+=-0.6$.
	The area A (B) is kinematically forbidden by
	$m_{\tilde\chi^0_2}+m_{\tilde\chi^0_3}>\sqrt{s}$
	$(m_{\tilde\ell_R}>m_{\tilde\chi^0_3})$.
	The gray  area is excluded by $m_{\tilde\chi_1^{\pm}}<104 $ GeV.
\label{plots_23}}
\end{figure}

The production of the neutralino pair 
$ e^+e^-\to\tilde\chi^0_2 \tilde\chi^0_3$
could make it easier to reconstruct the production plane
because both neutralinos decay. This
allows one to determine also asymmetry ${\mathcal A}_{I}$, 
which is a factor 2-3 larger than ${\mathcal A}_{II}$. 
We discuss the decay of the heavier neutralino 
$\tilde\chi^0_3$, which has a larger kinematically allowed region
in the $|\mu|$--$M_2$ plane than that of $\tilde\chi^0_2$. 
In Fig.~\ref{plots_23} we display the production cross section
$\sigma(e^+e^-\to\tilde\chi^0_2\tilde\chi^0_3 )$ which reaches 
100 fb. The cross section
$\sigma(e^+e^-\to\tilde\chi^0_2\tilde\chi^0_3 ) \times
BR(\tilde\chi^0_3\to\tilde\ell_R\ell_1)\times
BR(\tilde\ell_R\to\tilde\chi^0_1\ell_2)$
%{\bf
with BR($ \tilde\ell_R \to\tilde\chi^0_1\ell_2$) = 1
%}
is shown in Fig.~\ref{plots_23}b.
The asymmetry ${\mathcal A}_{II}$ is shown in Fig.~\ref{plots_23}d.
%{\bf
As to the $\varphi_{\mu}$ dependence of ${\mathcal A}_{I}$,
we found that 
for $\varphi_{\mu}=0.5\pi (0.1\pi)$ and $\varphi_{M_1}=0$,
$|{\mathcal A}_{I}|$ can reach 25\% (2\%) in the 
$M_2$--$|\mu|$ plane.
%}

\subsection{Energy distributions of the leptons
	  \label{Energy distributions of the leptons}}

In order to measure the asymmetries 
${\mathcal A}_{I}$~(\ref{TasymmetryI}) and 
${\mathcal A}_{II}$~(\ref{TasymmetryII}),
the two leptons $\ell_1$ and $\ell_2$ from  the neutralino 
(\ref{decay_1}) and slepton decay (\ref{decay_2})
have to be distinguished.
We therefore calculate the energy distributions of the leptons
from the first and second decay vertex 
%{\bf
in the laboratory system
(i.e. the cms of the incoming $e^+$ and $e^-$ beams).
%}
One can distinguish between the two leptons
event by event, if their energy distributions do not overlap.
If their energy distributions do overlap, only those leptons can 
be distinguished, whose energies are not both 
in the overlapping region.

The  energy distribution of lepton $\ell_1$ 
%{\bf
in the laboratory system
%}
has the form of a box with the endpoints:
 \begin{eqnarray}
  E_{\ell_1,min,max} &=&
  \frac{m_{\chi_i}^2-m_{\tilde\ell}^2}{2(E_{\chi_i} \pm q)},
\end{eqnarray}
with $q$ the neutralino momentum. 
The energy distribution 
%$\frac{1}{\sigma} \frac{d \sigma}{dE_8}$
of the second lepton  $\ell_2$ is obtained  
by integrating over the energy $E_{\tilde{\ell}}$ of the
propagating slepton: 
 \begin{eqnarray}
	 \frac{1}{\sigma} \frac{d \sigma}{dE_{\ell_2}}
	 &=& 
	 \frac{m_{\tilde{\ell}}^2~m_{\chi_i}^2}
	 {q[m_{\chi_i}^2-m_{\tilde{\ell}}^2]
		 [m_{\tilde{\ell}}^2-m_{\chi_1}^2]}  \times
 \left\{ \begin{array}{c@{\quad;\quad}c@{\quad\leq E_{\ell_2}\leq\quad}c}
 ln  \frac{\displaystyle E_{\ell_2}}{\displaystyle A}   &A&a \\
 ln  \frac{\displaystyle   a}{\displaystyle A}   &a&b \\
 ln  \frac{\displaystyle   B}{\displaystyle E_{\ell_2}} &b&B \\
 \end{array}  
 \right.
 \end{eqnarray}
with:
 \begin{eqnarray}
 A,B &=& \frac{m_{\tilde{\ell}}^2-m_{\chi_1}^2}
              {2m_{\tilde{\ell}}^2}
        \left( E_{\tilde{\ell},max} \mp 
                 \sqrt{E_{\tilde{\ell},max}^2-m_{\tilde{\ell}}^2} 
        \right)        \\
 a,b &=& \frac{m_{\tilde{\ell}}^2-m_{\chi_1}^2}
              {2m_{\tilde{\ell}}^2}
        \left( E_{\tilde{\ell},min} \mp 
                 \sqrt{E_{\tilde{\ell},min}^2-m_{\tilde{\ell}}^2} 
        \right)\\
    E_{\tilde{\ell},max,min} &=& 
             \frac{E_{\chi_i}(m_{\chi_i}^2+m_{\tilde{\ell}}^2) 
              \pm (m_{\chi_i}^2-m_{\tilde{\ell}}^2)\sqrt{E_{\chi_i}^2-m_{\chi_i}^2}}
                      {2 m_{\chi_i}^2}.
 \end{eqnarray}

\begin{figure}[h]
 \begin{picture}(20,20)(0,0)
\put(-1,17){\includegraphics{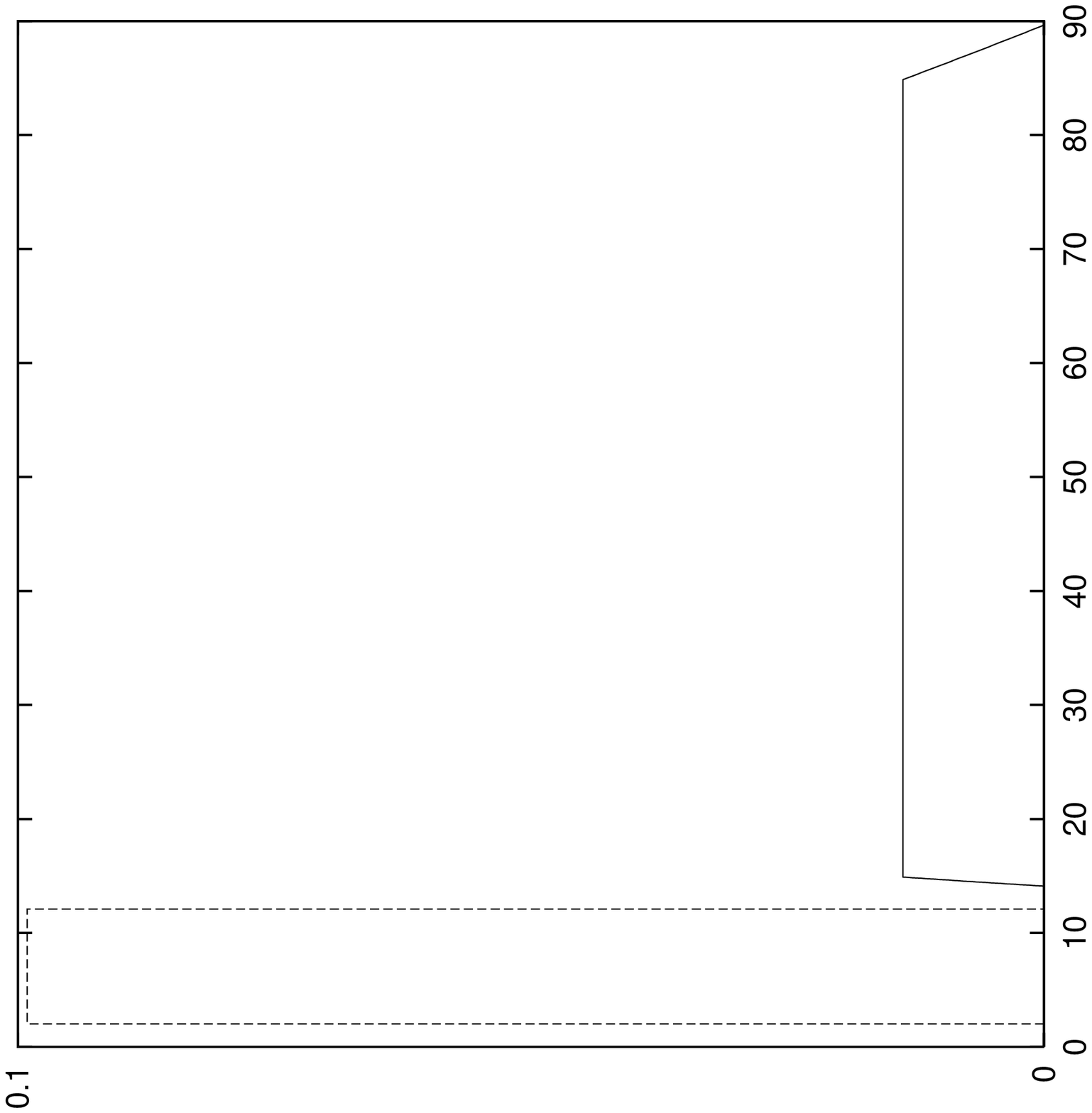}}
   \put(5.5,8.8){$E$~/GeV}
	\put(0,16.6){$ \frac{1}{\sigma}\frac{d\sigma}{dE}$}
\put(1.5,13.3){$\ell_1$}
\put(4.,10.7){$\ell_2$}
\put(0.5,8.8){Fig.~\ref{plotedist}a}
   \put(7,17){\includegraphics{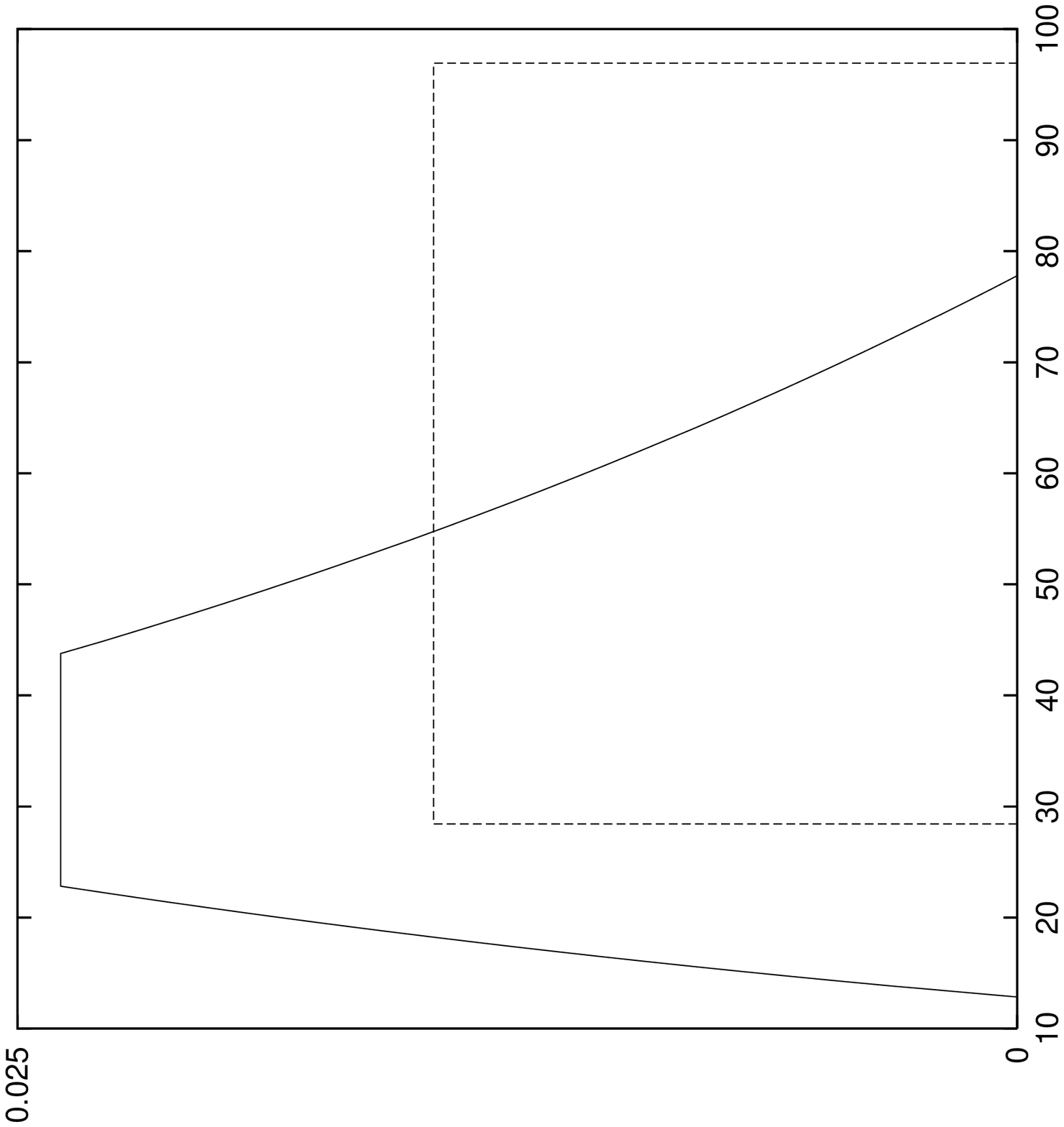}}
   \put(13.5,8.8){$E$~/GeV}
   \put(8,16.6){$ \frac{1}{\sigma}\frac{d\sigma}{dE} $ }
\put(10.3,15.3){$\ell_2$}
\put(12.5,13.6){$\ell_1$}
\put(8.5,8.8){Fig.~\ref{plotedist}b}
	\put(3,8){\includegraphics{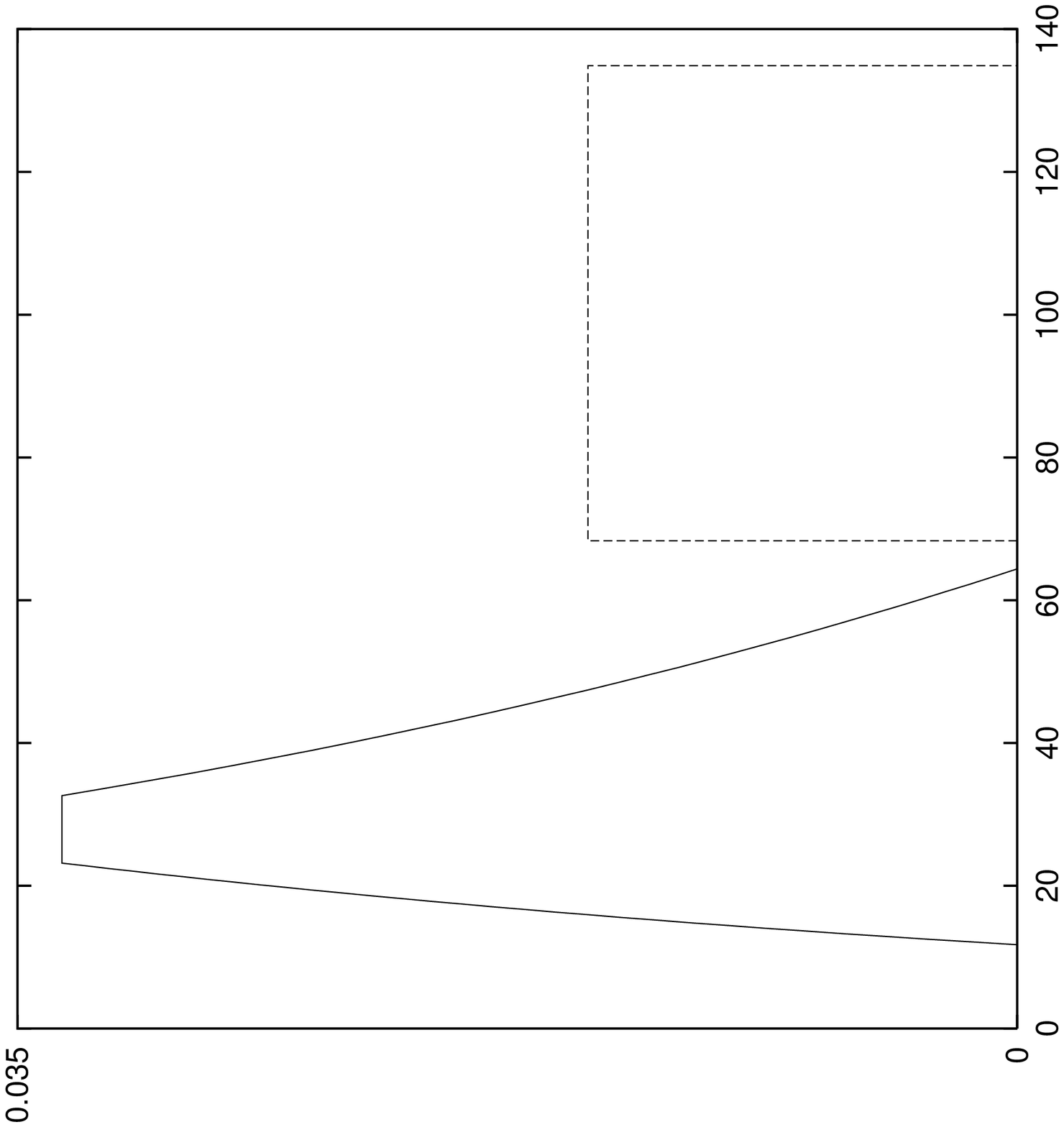}}
	\put(9.5,-0.3){$E$~/GeV}
	\put(4,7.6){$  \frac{1}{\sigma}\frac{d\sigma}{dE}$ }
	\put(5.8,6.3){$ \ell_2 $}
	\put(9.2,3.5){$\ell_1 $}
	\put(4.5,-0.3){Fig.~\ref{plotedist}c}

 \end{picture}
%\vspace*{.5cm}
\caption{
Possible types of energy distributions 
in the laboratory system
for $\ell_1$ (dashed line) and $\ell_2$ (solid line)
for $e^+  e^-\to \tilde{\chi}^0_1\tilde{\chi}^0_2$
and the subsequent decays $\tilde{\chi}^0_2\to \tilde{\ell}_R \ell_1$ 
and $\tilde{\ell}_R\to\tilde{\chi}^0_1\ell_2$,
for $M_2=300$ GeV, $m_{\tilde\ell_R}=180$ GeV,
$\tan\beta=10$ and $\{|\mu|,m_{\chi_1},m_{\chi_2}\}/{\rm GeV}=
\{200,140,185\},\;
\{300,145,240\},\;
\{500,150,300\}$ in a, b, c respectively.
\label{plotedist}}
\end{figure}

We show in Figs.~\ref{plotedist}a - c an example of the energy 
distributions of lepton $\ell_1$ (dashed line), and lepton $\ell_2$
(solid line), $\ell=e,\mu$, for
$e^+  e^-\to \tilde{\chi}^0_1\tilde{\chi}^0_2$
and the subsequent decays 
$\tilde{\chi}^0_2\to \tilde{\ell} \ell_1$ 
and $\tilde{\ell}\to\tilde{\chi}^0_1\ell_2$,
for $\tan \beta= 10$, $M_2=300$ GeV, $\varphi_{\mu}=0$ 
and $\varphi_{M_1}=0.5 \pi$ for  $|\mu|=200$, 300 and 500 GeV,
respectively. The parameters are chosen such that the
slepton mass $m_{\tilde{\ell}_R}=180$ GeV is constant, the
LSP mass $m_{\chi_1}=140,145,150 $ GeV is almost constant
whereas the neutralino mass $m_{\chi_2}=185,240,300$ GeV is increasing.
The mass difference between $\tilde{\ell}_R$ and $\tilde{\chi}^0_1$  
decreases ($\Delta m=40,35,30$ GeV), whereas
the mass difference between $\tilde{\chi}^0_2$ and $\tilde{\ell}_R$
increases ($\Delta m=5,60,120$ GeV).
The endpoints of the energy distributions of the decay leptons 
depend on  these mass differences. Thus, in 
Fig.~\ref{plotedist}a, the second lepton is more energetic 
than the first lepton. The energy distributions 
do not overlap and thus the two leptons can be distinguished
by measuring their energies.
This also holds for Fig.~\ref{plotedist}c, where 
the first  lepton is more energetic than the second one.
In Fig.~\ref{plotedist}b the two distributions overlap 
because the mass differences between $\tilde\chi^0_1$, $\tilde\ell_R$
and $\tilde\chi^0_2$ are similar. One has to apply cuts in order 
to distinguish the two leptons, which reduce the number of events. 

A potentially large background may be due to slepton production
$ e^+e^-\to\tilde\ell^+ \tilde\ell^-
\to\ell^+\ell^-\tilde\chi^0_1\tilde\chi^0_1$.
However, these reactions would lead generally to ''two-sided
events``, whereas the events from
$ e^+e^-\to\tilde\chi^0_1 \tilde\chi^0_i
\to\ell^+\ell^-\tilde\chi^0_1\tilde\chi^0_1$
are ''one-sided events``. Moreover, the background
reaction is CP-even and will not give rise to a CP asymmetry, because
the sleptons are scalars and their decay is a two-body one.

\section{Summary and conclusion
	\label{Summary and conclusion}}

We have considered two CP sensitive triple-product asymmetries in 
neutralino production
$e^+e^- \to\tilde\chi^0_i \tilde\chi^0_j$
and the subsequent leptonic two-body decay chain of one  neutralino
$\tilde\chi^0_i \to \tilde\ell  \ell$,
$ \tilde\ell \to \tilde\chi^0_1  \ell$ for
$ \ell= e,\mu,\tau$. The CP sensitive contributions to the asymmetries
are present  already at tree level and are due to spin effects in the 
production process of two different neutralinos. 
The asymmetries are induced only if 
CP-violating phases of the 
gaugino and higgsino mass parameters $M_1$ and/or $\mu$ are present 
in the neutralino sector of the MSSM.

In a numerical study for 
$e^+e^- \to\tilde\chi^0_1 \tilde\chi^0_2$
and neutralino decay into a right slepton
$\tilde\chi^0_2 \to \tilde\ell_R \ell$
we have shown that the asymmetries can be 
as large as 25\%. 
They can be sizeable even for a small phase of $\mu$,
which is suggested by the experimental limits on EDMs. 
The asymmetries are simlar for the processes 
$e^+e^- \to\tilde\chi^0_1 \tilde\chi^0_3$ and
$e^+e^- \to\tilde\chi^0_2 \tilde\chi^0_3$.
Depending on the MSSM scenario, 
our proposed asymmetries should be accessible in future  
electron-positron linear collider experiments in the
500 GeV range. Longitudinally polarized electron and positron
beams can considerably enhance both asymmetries
and production cross sections.

\section{Acknowledgement}

%{\bf
We thank S. Hesselbach and T. Kernreiter for useful discussions.
%}
This work was supported by the `Fonds zur
F\"orderung der wissenschaftlichen Forschung' (FWF) of Austria, projects
No. P13139-PHY and No. P16592-N02 and by the European Community's
Human Potential Programme under contract HPRN-CT-2000-00149.
This work was also supported by the 'Deutsche Forschungsgemeinschaft'
(DFG) under contract Fr 1064/5-1.
O.K. acknowledges support from the
\emph{Bayerische Julius-Maximilians Universit\"at W\"urzburg}.

\newpage

\begin{appendix}
\noindent{\Large\bf Appendix}

\setcounter{equation}{0}
\renewcommand{\thesubsection}{\Alph{section}.\arabic{subsection}}
\renewcommand{\theequation}{\Alph{section}.\arabic{equation}}
\section{Momentum and spin vectors
     \label{Representation of momentum and spin vectors}}
\setcounter{equation}{0}

We choose a coordinate frame in the center of
mass system such that the momentum of neutralino
$\tilde \chi ^0_j$ points in the $z$-direction. 
The scattering  angle is 
$\theta \angle (\vec p_{e^-},\vec p_{\tilde \chi_j})$ and 
the azimuth $\phi$ can be chosen to zero. The momenta are given by: 
   \begin{eqnarray}
  && p_{e^-} = E_b(1,-\sin\theta,0, \cos\theta),\quad
     p_{e^+} = E_b(1, \sin\theta,0,-\cos\theta),\\
 &&  p_{\chi_i} = (E_{\chi_i},0,0,-q),\quad
     p_{\chi_j} = (E_{\chi_j},0,0, q),
   \end{eqnarray}
with the beam energy $E_b=\sqrt{s}/2$ and
\begin{eqnarray}
 &&   E_{\chi_i} =\frac{s+m_{\chi_i}^2-m_{\chi_j}^2}{2 \sqrt{s}},\quad
    E_{\chi_j} =\frac{s+m_{\chi_j}^2-m_{\chi_i}^2}{2 \sqrt{s}},\quad
      q =\frac{\lambda^{\frac{1}{2}}
             (s,m_{\chi_i}^2,m_{\chi_j}^2)}{2 \sqrt{s}}, 
\end{eqnarray}
with $m_{\chi_i}, m_{\chi_j}$ the masses of the neutralinos and 
$\lambda(x,y,z) = x^2+y^2+z^2-2xy-2xz-2yz$.
The  three spin basis vectors of $\tilde{\chi}^0_i$ are chosen by:
\begin{eqnarray}
	&&  s^1_{\chi_i}=(0,-1,0,0),\quad
    s^2_{\chi_i}=(0,0,1,0),\quad
    s^3_{\chi_i}=\frac{1}{m_{\chi_i}}(q,0,0,-E_{\chi_i}).
\end{eqnarray} 
Together with the unit 
momentum vector $p_{\chi_i}/m_{\chi_i}$ of $\tilde{\chi}^0_i$, the spin 
basis vectors form an orthonormal set.
The momenta and energies of the leptons are:
 \begin{eqnarray}
&&   p_{{\ell}_1} = (                        E_{{\ell}_1},
            -|\vec p_{{\ell}_1}| \sin \theta_{1} \cos \phi_{1},
             |\vec p_{{\ell}_1}| \sin \theta_{1} \sin \phi_{1},
				 -|\vec p_{{\ell}_1}| \cos \theta_{1}), \\
&& p_{{\ell}_2} = (                        E_{{\ell}_2},
            -|\vec p_{{\ell}_2}| \sin \theta_{2} \cos \phi_{2},
             |\vec p_{{\ell}_2}| \sin \theta_{2} \sin \phi_{2},
				-|\vec p_{{\ell}_2}| \cos \theta_{2}),\\
&& E_{{\ell}_1} = |\vec p_{{\ell}_1}| = 
	 \frac{m_{\chi_i}^2-m_{\tilde{{\ell}}}^2}{2(E_{\chi_i}-q \cos
			 \theta_{1})}, \quad
	E_{{\ell}_2} = |\vec p_{{\ell}_2}| =
	  \frac{m_{\tilde{{\ell}}}^2-m_{\chi_1}^2 }
	  {2(E_{\tilde {\ell} }-|\vec p_{\chi_i}-\vec p_{{\ell}_1}| \cos\theta_{D_2})},
	 	   \label{energy6}
 \end{eqnarray}
with $\theta_{1}=\angle (\vec p_{\ell_{1}} ,\vec p_{\chi_i})$,
 $\theta_{2}=\angle (\vec p_{\ell_{2}} ,\vec p_{\chi_i})$
and the decay angles 
$\theta_{D_2} \angle (\vec p_{\tilde {\ell} },\vec p_{{\ell}_2})$
and $\theta_{D_1} \angle (\vec p_{\chi_i},\vec p_{\tilde {\ell}})$
(see Fig.~\ref{shematic picture}):
 \begin{equation}
\cos\theta_{D_2}=\cos\theta_{D_1}\cos\theta_{2}-
\sin\theta_{D_1}\sin\theta_{2}\cos(\phi_{2}-\phi_{1}), \quad
\cos\theta_{D_1}=\frac{\vec p_{\chi_i } (\vec p_{\chi_i}-\vec p_{{\ell}_1})}
{ |\vec p_{\chi_i }|~ | \vec p_{\chi_i}-\vec p_{{\ell}_1} | }.
  \end{equation}

\section{Phase space
     \label{Phase space}}
\setcounter{equation}{0}

The Lorentz invariant phase space element for the neutralino 
production (\ref{production}) and the decay 
chain (\ref{decay_1})-(\ref{decay_2}) can be decomposed
into the two-body  phase space elements:
\begin{eqnarray}
 &&d{\rm Lips}(s,p_{\chi_j },p_{{\ell}_1},p_{\chi_1},p_{{\ell}_2}) =
	 \nonumber \\ 
&&\frac{1}{(2\pi)^2}~d{\rm Lips}(s,p_{\chi_i},p_{\chi_j} )
 ~d s_{\chi_i} ~d{\rm Lips}(s_{\chi_i},p_{\tilde {\ell}},p_{{\ell}_1})
 ~d s_{\tilde {\ell}}~d{\rm Lips}(s_{\tilde {\ell}},p_{\chi_1},p_{{\ell}_2}),\label{Lips}
 \end{eqnarray}
\begin{eqnarray}
	d{\rm Lips}(p_{\chi_i },p_{\chi_j })&=&
	\frac{q}{8\pi\sqrt{s}}~\sin\theta~ d\theta, \\
	d{\rm Lips}(s_{\chi_i},p_{\tilde {\ell}},p_{{\ell}_1})&=&
	\frac{1}{2(2\pi)^2}~
	\frac{|\vec p_{{\ell}_1}|^2}{m_{\chi_i}^2-m_{\tilde {\ell}}^2}
	~d\Omega_1,\\
	d{\rm Lips}(s_{\tilde {\ell}},p_{\chi_1},p_{{\ell}_2})&=&
\frac{1}{2(2\pi)^2}~\frac{|\vec p_{{\ell}_2}|^2}{m_{\tilde {\ell}}^2-m_{\chi_1}^2}
	~d\Omega_2,
\end{eqnarray}
with $s_{\chi_i}=p^2_{\chi_i}$, $s_{\tilde {\ell}}=p^2_{\tilde {\ell}}$ and 
$ d\Omega_i=\sin\theta_i~ d\theta_i~ d\phi_i$.
  
\section{Neutralino production and decay matrices
  \label{Neutralino production and decay matrices}}
\setcounter{equation}{0}

In this Section we give the analytical expressions for
$P,\Sigma_P^2$ and $D_1,D_2,\Sigma_{D1}^a$ in the center of mass system. 
Expressions for $\Sigma_P^{1,3}$ can be found in \cite{gudi1}.

\subsection{Neutralino production 
     \label{Neutralino production}}

The analytic expression $P$ of Eq.~(\ref{eq_3}) is independent of the 
neutralino polarization. It can be  decomposed into 
contributions from the different production channels \cite{gudi1}:
\begin{equation}
P=P(Z Z)+P(Z \tilde{e}_R)+P(Z \tilde{e}_L)+
P(\tilde{e}_R \tilde{e}_R)+P(\tilde{e}_L \tilde{e}_L),\label{eq_15}
\end{equation}
with
\begin{eqnarray}
P(Z Z)&=&
2 \frac{g^4}{\cos^4\theta_W}|\Delta^s(Z)|^2
[(1-P^3_-P^3_+)(L_{e}^2+R_{e}^2) - (P^3_- - P^3_+)(L_{e}^2-R_{e}^2)]
E_b^2 
\nonumber\\& &
\Big\{ |O^{''R}_{ij}|^2 (E_{\chi_i} E_{\chi_j}+q^2\cos^2\theta)-[(Re O^{''R}_{ij})^2 
-(Im O^{''R}_{ij})^2]
m_{\chi_i} m_{\chi_j}\Big\}\label{eq_16},\label{P_1}\\
P(Z \tilde{e}_R)&=&
\frac{g^4}{\cos^2 \theta_W} 
R_{e} (1+P^3_{-})(1-P_{+}^3) E_b^2
Re\Big\{\Delta^s(Z)
\nonumber\\& & 
\Big[
-(\Delta^{t*}(\tilde{e}_R) f^{R*}_{e i} 
f^{R}_{e j} O^{''R*}_{ij}
+\Delta^{u*}(\tilde{e}_R) f^R_{e i} f^{R*}_{e j}
O^{''R}_{ij}) m_{\chi_i} m_{\chi_j}
\nonumber\\& & 
-(\Delta^{t*}(\tilde{e}_R) f^{R*}_{e i} f^{R}_{e j} 
O^{''R}_{ij}
-\Delta^{u*}(\tilde{e}_R) f^R_{e i} f^{R*}_{e j} 
O^{''R*}_{ij})
2 E_b q \cos\theta
\nonumber\\
& &
+(\Delta^{t*}(\tilde{e}_R) f^{R*}_{e i} f^{R}_{e j}
 O^{''R}_{ij}
+\Delta^{u*}(\tilde{e}_R) f^R_{e i} f^{R*}_{e j} 
O^{''R*}_{ij})
(E_{\chi_i} E_{\chi_j}+q^2\cos^2\theta)
\Big]\Big\},\label{P_2}\\
P(\tilde{e}_R \tilde{e}_R)&=&
\frac{g^4}{4} (1+P^3_{-})(1-P_{+}^3)E_b^2 
\Big\{|f^R_{e i}|^2 |f_{e j}^R|^2 \nonumber\\
& &\mbox{\hspace*{-.5cm}}\Big[ (|\Delta^t(\tilde{e}_R)|^2
+|\Delta^u(\tilde{e}_R)|^2)
(E_{\chi_i} E_{\chi_j}+q^2 \cos^2\theta)
-(|\Delta^t(\tilde{e}_R)|^2-|\Delta^u(\tilde{e}_R)|^2)
2 E_b q \cos\theta\Big]
\nonumber\\
& &\mbox{\hspace*{-.5cm}}
-Re\{(f^{R*}_{e i})^2 (f^R_{e j})^2
     \Delta^u(\tilde{e}_R)\Delta^{t*}(\tilde{e}_R)\}
2 m_{\chi_i} m_{\chi_j}\Big\}.\label{P_3}
\end{eqnarray}
To obtain the quantities
$P(Z\tilde{e}_L),P(\tilde{e}_L \tilde{e}_L)$ one
has to exchange in
Eqs.~(\ref{P_1})~-~(\ref{P_3})
\begin{eqnarray}\nonumber
     &&\Delta^{t}(\tilde{e}_R)\to\Delta^{t}(\tilde{e}_L),\quad
	  \Delta^{u}(\tilde{e}_R)\to\Delta^{u}(\tilde{e}_L),\quad
	  P^3_{-}\to P^3_{+},\quad P^3_{+}\to P^3_{-}\\
	 &&R_{e}\to L_{e},\quad
         O^{''R}_{ij}\to O^{''L}_{ij},\quad
           f_{e i}^R\to f_{e i}^L,\quad
			  f_{e j}^R\to f_{e j}^L. \label{exchange}
\end{eqnarray}
The propagators are defined as follows:
    \begin{equation}
        \Delta^s(Z)=\frac{i}{s-m^2_Z+im_Z\Gamma_Z},\quad
                  \Delta^{t}(\tilde{e}_{R,L})=
            \frac{i}{t-m^2_{\tilde{e}_{R,L}}},\quad
              \Delta^u (\tilde{e}_{R,L})=
             \frac{i}{u-m^2_{\tilde{e}_{R,L}}},\label{eq_11}
         \end{equation}
where $m$ and $\Gamma$ denote the mass and width of the exchanged
particle, respectively, and $s=(p_{e^-}+p_{e^+})^2$, $t=(p_{e^-}-p_{\chi_j})^2$, and $u=(p_{e^-}-p_{\chi_i})^2$.
The longitudinal beam polarization of $e^{-} (e^{+})$ are denoted by
$P_{-}^3 (P_{+}^3)$, respectively.
Generally the contributions from the exchange  
of $\tilde{e}_{R}$ ($\tilde{e}_{L}$) selectron exchange
is enhanced and that of $\tilde{e}_{L}$ ($\tilde{e}_{R}$) is
suppressed for $P_-^3>0,P_+^3<0~(P_-^3<0,P_+^3>0)$.

\subsection{Neutralino polarization 
     \label{Neutralino polarization}}

The analytic expressions for the 
coefficient $\Sigma^2_P$ in Eq.~(\ref{rhoP}) which describes the 
transversal polarization of neutralino $\tilde{\chi}^0_i$ perpendicular 
to the production plane decomposes into 
contributions from the different production channels \cite{gudi1}:
   \begin{equation}
     \Sigma_P^a(\tilde{\chi}^0_i)=
     \Sigma_P^a(\tilde{\chi}^0_i,ZZ)
   + \Sigma_P^a(\tilde{\chi}^0_i,Z\tilde{e}_R)
   + \Sigma_P^a(\tilde{\chi}^0_i,Z\tilde{e}_L)
   + \Sigma_P^a(\tilde{\chi}^0_i,\tilde{e}_R \tilde{e}_R)
   + \Sigma_P^a(\tilde{\chi}^0_i,\tilde{e}_L \tilde{e}_L).\label{eq_27}
\end{equation}
In the center of mass system they read \cite{gudi1}:
   \begin{eqnarray}
     \Sigma_P^2 \Big{(} \tilde{\chi}^0_i,ZZ)&=&
            -4 (\frac{g^2}{\cos^2\theta_W} \Big) ^2 |\Delta^s(Z)|^2  
				[(1-P^3_-P^3_+)(L_{e}^2-R_{e}^2) - 
				(P^3_- - P^3_+)(L_{e}^2+R_{e}^2)] \nonumber\\
                 & & \label{s2_1} \times \, m_{\chi_j} q E_b^2 
                 \sin\theta Re(O^{''R}_{ij}) Im(O^{''R}_{ij}),\\
    \Sigma_P^2(\tilde{\chi}^0_i,Z \tilde{e}_R) &=& 
	 \frac{g^4}{\cos^2\theta_W} R_{e} (1+P^3_{-})(1-P_{+}^3) m_{\chi_j}
          E_b^2 q \sin\theta \nonumber\\
          & & \times \, Im\Big\{\Delta^s(Z)
            \big[f^{R}_{e i}f^{R*}_{e j} O^{''R}_{ij} 
           \Delta^{u*}(\tilde{e}_R) -f^{R*}_{e i}f^{R}_{e j}
            O^{''R*}_{ij} \Delta^{t*}(\tilde{e}_R) \big]\Big\},\nonumber\\ 
          & &\label{s2_2}\\
     \Sigma_P^2(\tilde{\chi}^0_i,\tilde{e}_R \tilde{e}_R)&=& 
	  -\frac{g^4}{2} (1+P^3_{-})(1-P_{+}^3)  m_{\chi_j} E_b^2 q \sin\theta 
	  \nonumber\\ & & \times \,Im\Big\{(f^{R*}_{e i})^2 (f^{R}_{e j})^2
          \Delta^u (\tilde{e}_R)\Delta^{t*}
			 (\tilde{e}_R)\Big\}.\label{s2_3}
     \end{eqnarray}
To obtain the expressions for 
$\Sigma_P^2(\tilde{\chi}^0_i,Z \tilde{e}_L)$ and
$\Sigma_P^2(\tilde{\chi}^0_i,\tilde{e}_L \tilde{e}_L)$
one has to apply the exchanges (\ref{exchange})
in Eq.~(\ref{s2_1})~-~(\ref{s2_3}).

\subsection{Neutralino decay matrix
     \label{Neutralino decay matrix}}

The neutralino decay matrix is given by Eq.~(\ref{eq_3}):
   \begin{eqnarray} \nonumber
		\rho_{D1}(\tilde \chi_i^0)_{\lambda_i' \lambda_i} & = & 
           \delta_{\lambda_i' \lambda_i} D_1 +
			  \sum_a \sigma^a_{\lambda_i' \lambda_i}\Sigma^a_{D1}.
    \end{eqnarray}
The expansion coefficients $D_1$ and $\Sigma^a_{D1}$ for
the decay into the right slepton are ($\ell=e,\mu$):
\begin{eqnarray}\label{D_1}
      D_1 & = & \frac{g^2}{2} |f^{R}_{\ell i}|^2 (m_{\chi_i}^2 -m_{\tilde{\ell}}^2 ),\\
		\Sigma^a_{D1} &=&   g^2 |f^{R}_{\ell i}|^2 m_{\chi_i} (s^a_{\chi_i}
			\cdot p_{\ell_1}),\label{SigmaD_1}
   \end{eqnarray}
respectively, with $s^a_{\chi_i}$ the neutralino spinvector and $p_{\ell_1}$ 
the lepton $\ell_1$ momentum vector.
For the decay into the left selectron or smuon they are:
   \begin{eqnarray}
      D_1 & = & \frac{g^2}{2} |f^{L}_{\ell i}|^2 (m_{\chi_i}^2 -
			m_{\tilde{\ell}}^2 ),\\
		\Sigma^a_{D1} &=&  - g^2 |f^{L}_{\ell i}|^2 m_{\chi_i} (s^a_{\chi_i} \cdot p_{\ell_1}).
   \end{eqnarray}
For the decay into the stau $\tilde \tau_k \; (k=1,2)$ they read:
   \begin{eqnarray}\label{plusterm}
		D_1 & = & \frac{g^2}{2} (
			|a_{ki}^{\tilde \tau}|^2 +|b_{ki}^{\tilde \tau}|^2 )
				(m_{\chi_i}^2 - m_{\tilde{\tau}_k}^2 ),\\
				\Sigma^a_{D1} &=&   g^2 (
			|a_{ki}^{\tilde \tau}|^2-|b_{ki}^{\tilde \tau}|^2 )
			m_{\chi_i}	(s^a_{\chi_i} \cdot p_{\ell_1}\label{minusterm}).
   \end{eqnarray}
The decay of the right ($R$) or left ($L$) slepton into lepton 
and $\tilde \chi_1^0 $ is given by ($\ell=e,\mu$):
\begin{eqnarray}
	D_2 & = & g^2 |f^{R,L}_{\ell 1}|^2 
	( m_{\tilde{\ell}}^2-m_{\chi_1}^2 ),
\end{eqnarray}
and into staus $\tilde\tau_k$ by:
\begin{eqnarray}
			D_2 & = &g^2 (
			|a_{k1}^{\tilde \tau}|^2 +|b_{k1}^{\tilde \tau}|^2 )
				(m_{\tilde{\tau}_k}^2-m_{\chi_1}^2 ).
		\end{eqnarray}

\end{appendix}

\end{document}